\documentclass[a4paper,11pt]{article}
\pdfoutput=1
\usepackage{jcappub}
\usepackage{bm}
\usepackage{latexsym}
\usepackage{dcolumn}
\usepackage{amsfonts,amssymb,amsmath}
\usepackage{graphicx,epsfig}
\usepackage{psfrag}
\usepackage{soul}
\usepackage[normalem]{ulem}
\usepackage{subfigure}
\usepackage[dvipsnames]{xcolor}
\usepackage{rotating}
\usepackage{hyperref}
\usepackage{tikz}
\usepackage{soul}
\usepackage[utf8]{inputenc}
\hypersetup{
	bookmarks=true,         
	unicode=false,          
	pdftoolbar=true,        
	pdfmenubar=true,        
	pdffitwindow=false,     
	pdfstartview={FitH},    
	pdftitle={My title},    
	pdfauthor={Author},     
	pdfsubject={Subject},   
	pdfcreator={Creator},   
	pdfproducer={Producer}, 
	pdfkeywords={keyword1} {key2} {key3}, 
	pdfnewwindow=true,      
	colorlinks=true,       
	linkcolor=red,          
	citecolor=cyan,        
	filecolor=magenta,      
	urlcolor=blue,           
	linktocpage=true
}

\newcommand{\dd}{\mathrm{d}}
\newcommand{\ee}{\mathrm{e}}

\def\beq{\begin{equation}}
\def\efq{\end{equation}}
\def\br{\begin{eqnarray}}
\def\er{\end{eqnarray}}
\def\benu{\begin{enumerate}}
	\def\efnu{\end{enumerate}}

\def\be{\begin{equation}}
\def\ee{\end{equation}}
\def\ba{\begin{eqnarray}}
\def\ea{\end{eqnarray}}

\DeclareUnicodeCharacter{2212}{-}

\begin{document}
	\title{Probing Early Modification of Gravity with {\it Planck}, ACT and SPT}
	
\author[a]{Guillermo Franco Abellán,}
\author[b,c]{Matteo Braglia,}
\author[d,e,c]{Mario Ballardini,}
\author[c,f]{Fabio Finelli}
\author[g]{and Vivian Poulin}

\affiliation[a]{GRAPPA Institute, Institute for Theoretical Physics Amsterdam, University of Amsterdam, Science Park 904, 1098 XH Amsterdam, The Netherlands}

\affiliation[b]{Center for Cosmology and Particle Physics, New York University, 726 Broadway, New York, NY 10003, USA}

\affiliation[c]{INAF/OAS Bologna, via Piero Gobetti 101, 40129 Bologna, Italy}

\affiliation[d]{Dipartimento di Fisica e Scienze della Terra, Universit\`a degli Studi di Ferrara, via Giuseppe Saragat 1, 44122 Ferrara, Italy}

\affiliation[e]{INFN, Sezione di Ferrara, via Giuseppe Saragat 1, 44122 Ferrara, Italy}

\affiliation[f]{INFN, Sezione di Bologna, viale C. Berti Pichat 6/2, 40127 Bologna, Italy}

\affiliation[g]{Laboratoire Univers \& Particules de Montpellier (LUPM), CNRS \& Universit\'e de Montpellier (UMR-5299), Place Eug\`ene Bataillon, F-34095 Montpellier Cedex 05, France}
\emailAdd{g.francoabellan@uva.nl}
\emailAdd{mb9289@nyu.edu}
\emailAdd{mario.ballardini@unife.it}
\emailAdd{fabio.finelli@inaf.it}
\emailAdd{vivian.poulin@umontpellier.fr}

 \abstract{
 We consider a model of early modified gravity (EMG) that was recently proposed as a candidate to resolve the Hubble tension. The model consists of a scalar field $\sigma$ with a nonminimal coupling (NMC) to the Ricci curvature of the form $F(\sigma) = M_{\mathrm{pl}}^2+\xi\sigma^2$ and an effective mass induced by a quartic potential $V(\sigma) = \lambda \sigma^4/4$. We present the first analyses of the EMG model in light of the latest ACT DR4 and SPT-3G data in combination with full \emph{Planck} data, and find a $\gtrsim 2\sigma$ preference for a non-zero EMG contribution from a combination of primary CMB data alone, mostly driven by ACT-DR4 data. This is different from popular `Early Dark Energy' models, which are detected only when the high-$\ell$ information from $\emph{Planck}$ temperature is removed. We find that the NMC plays a key role in controlling the evolution of density perturbations that is favored by the data over the minimally coupled case.  Including measurements of supernovae luminosity distance from Pantheon+, baryonic acoustic oscillations and growth factor from BOSS, and CMB lensing of {\em Planck} leaves the preference unaffected. In the EMG model, the tension with S$H_0$ES is alleviated from $\sim 6\sigma$ to $\sim 3\sigma$. Further adding S$H_0$ES data raises the detection of the EMG model above $5\sigma$.
 }
	\maketitle
\section{Introduction}
The precise determination of the Hubble constant, denoted by $H_0$, has emerged as a formidable challenge in modern cosmology \cite{Knox:2019rjx,DiValentino:2020zio,DiValentino:2021izs, Schoneberg:2021qvd,Shah:2021onj,Abdalla:2022yfr}. The Hubble tension, a persistent discrepancy between direct and indirect measurements of the Universe's expansion rate \cite{Verde:2019ivm}, has ignited intense scrutiny within the field. In particular, the inferred value of $H_0$ in the standard flat $\Lambda$CDM cosmological model using cosmic microwave background (CMB) measurements from the {\em Planck} DR3 \cite{Planck:2018vyg}, i.e. $H_0 = 67.36 \pm 0.54\,{\rm km}\,{\rm s}^{-1}\,{\rm Mpc}^{-1}$, deviates by $5\sigma$ from the value obtained through the distance ladder calibration of type Ia supernovae (SN Ia) from the Pantheon+ compilation \cite{Carr:2021lcj} by the S$H_0$ES collaboration \cite{Riess:2021jrx}, i.e. $H_0 = 73.0 \pm 1.0\,{\rm km}\,{\rm s}^{-1}\,{\rm Mpc}^{-1}$ (see also Refs.~\cite{Freedman:2019jwv,Kamionkowski:2022pkx,Uddin:2023iob,Freedman:2023zdo} for discussions). Despite intense experimental efforts to better understand the systematic effects that could plague either measurements, this discrepancy has worsened since the first release of {\em Planck} data, triggering the need to explore modifications to the $\Lambda$CDM model, for instance via the addition of cosmological dynamical fields. In particular, it has been shown that resolving the Hubble tension most likely requires deviations away from $\Lambda$CDM in the pre-recombination era, so as to reduce the ``sound horizon'' $r_s$, the standard ruler that is used to calibrate the CMB and galaxy surveys and infer a value of the Hubble constant. One possible way to reduce $r_s$ is to briefly increase the Hubble rate around matter-radiation equality, for instance via a new form of Dark Energy active at early time \cite{Karwal:2016vyq,Poulin:2018cxd} dubbed `early dark energy' (EDE) (see  Refs.~\cite{Kamionkowski:2022pkx,Poulin:2023lkg} for reviews), or by modifying general relativity (GR).

Although GR stands as the standard theory of gravity used in cosmological models, alternatives such as scalar-tensor theories of gravity have been proposed as potential solutions to the Hubble tension. Even the simplest scalar-tensor theories of gravity such as Jordan-Brans-Dicke \cite{Jordan:1949zz,Brans:1961sx}, with a scalar field regulating the time evolution of the effective Newton constant, acts by injecting energy into the cosmic fluid at early times proportionally to the strength of the coupling to the Ricci scalar, thereby naturally causing a larger CMB-inferred value of $H_0$  \cite{Umilta:2015cta}.
Among the extensions of this archetypal model - including nonminimal coupled scalar fields
\cite{Rossi:2019lgt,Ballardini:2020iws,Braglia:2020iik,Ballesteros:2020sik,Adi:2020qqf,Benevento:2022cql,Kable:2023bsg}, different scalar field potentials \cite{Ballardini:2016cvy,SolaPeracaula:2019zsl}, sign of the kinetic term \cite{Ballardini:2023mzm}, and the addition of allowed Galileon-like terms within Horndeski theory \cite{Zumalacarregui:2020cjh,Ferrari:2023qnh}, or phenomenological parameterizations~\cite{Lin:2018nxe} - the most successful model studied so far has been the `early modified gravity' (EMG) model presented in Ref.~\cite{Braglia:2020auw}. In the EMG model, a self-interacting scalar field non-minimally coupled to gravity evolves from the radiation era, driven by the coupling to non-relativistic matter. Then, the scalar field enters in a coherent oscillations regime due to a self-interaction quartic term in the potential and its energy density dilutes faster than non-relativistic matter after recombination. The decrease in the amplitude of the scalar field provides a natural way to suppress modifications to gravity at late redshifts, leading to excellent agreement with laboratory and Solar System tests of GR without the need for screening mechanisms.

Although theoretical motivations and initial conditions differ, EMG and EDE models have in common a localized energy injection, induced by a non-negligible scalar field energy in the pre-recombination era that rapidly decays around matter-radiation equality. It has been found that they  perform nearly identically in alleviating the $H_0$ tension \cite{Schoneberg:2021qvd}.  In fact, given the use of a quartic term in the potential, the EMG model proposed in Ref.~\cite{Braglia:2020auw} reduces to the Rock'n'Roll (RnR) model  of EDE \cite{Agrawal:2019lmo} in the minimally coupled limit. In RnR a quartic term in the potential substitutes the cosine potentials introduced originally in the EDE \cite{Poulin:2018cxd} models. Yet, the RnR model does not fair as good in resolving the Hubble tension \cite{Smith:2019ihp}: this suggests that the nonminimal coupling plays a key role in resolving the Hubble tension that is yet to be understood.  

Another important aspect of the 
current status of CMB measurements lies in the differences between the 
 results obtained from the {\em Planck} satellite and the latest measurements from the Atacama Cosmology Telescope (ACT) ground-based experiment \cite{ACT:2020gnv}. ACTPol has recently produced CMB maps with higher angular resolution compared to those from the {\em Planck} satellite, offering precise information about the early universe and its subsequent evolution. Previous studies \cite{Hill:2021yec,Poulin:2021bjr,Smith:2022hwi} have demonstrated disparate outcomes for EDE when analysing {\em Planck} versus ACTPol data alone or in combination, with different $\ell$-cuts in temperature to reduce the potential correlations between the two data sets. In particular, when ACTPol is analysed in combination with a WMAP-like cut to {\em Planck} data, a tantalizing hint of EDE at $3\sigma$ was reported \cite{Hill:2021yec,Poulin:2021bjr,Smith:2022hwi} (with a value of $H_0$ compatible with direct measurements), but the hint is reduced when the full temperature data from {\em Planck} are included. In addition, the South Pole Telescope (SPT) collaboration recently released their latest CMB temperature and polarization power spectrum measurements \cite{SPT-3G:2022hvq}, with potentially strong implications models attempting to resolve the Hubble tension. 
Analyses of the EMG model in light of these new datasets is still lacking.

The aim of this paper is twofold. We first aim at providing a comprehensive investigation of the EMG fit to a compilation of the most up-to-date CMB data -- beyond what was done in Ref.~\cite{Braglia:2020auw} for {\em Planck} data only -- and  in a similar way to what has been done for EDE in Refs.~\cite{Hill:2021yec,Poulin:2021bjr,Smith:2022hwi}. 
In particular, we report a hint of a preference for EMG in analyses that include ACT data. We find that the NMC plays an essential role in setting the preference over simple power-law EDE models like RnR, that are excluded as a solution to the $H_0$ tension. Our second goal is therefore to develop a deeper understanding of the differences between EMG \cite{Braglia:2020auw} and the RnR EDE model proposed in \cite{Agrawal:2019lmo}.

The paper is organized as follows. In Sec.~\ref{sec:model_data} we review the basics of the EMG model and presents our analysis setup and the various datasets used in this analysis, while Sec.~\ref{sec:results} presents our main results. We focus first on analyses of EMG in light of CMB data in Sec.~\ref{sec:EMG_CMB}, investigate the role of the NMC in driving the preference in Sec.~\ref{sec:EMG_RnR}, and establish the residual level of tension with S$H_0$ES in Sec.~\ref{sec:EMG_Ext}. We finally conclude in Sec.~\ref{sec:concl}. To keep the paper self-contained, App.~\ref{app:eq} provides more details on the relevant equations driving the EMG dynamics in cosmological linear perturbation theory. App.~\ref{app:SW} expands upon the different signatures that the EMG and RnR models leave on the CMB spectra, in particular regarding the so-called ``Sachs-Wolfe'' contribution. We provide all relevant $\chi^2$ statistics, tension metrics and results of Monte Carlo Markov Chain (MCMC) in App.~\ref{app:chi2}. 

\section{Model and Data}
\label{sec:model_data}
\subsection{The EMG model}
\label{sec:model}

In this section, we briefly introduce the EMG model and its dynamics. We refer the reader to Ref.~\cite{Braglia:2020auw} for more details.
The action describing the model is that of a scalar field non-minimally coupled to the Ricci scalar $R$:
\begin{equation}
	\label{eq:model}
	S = \int \dd^{4}x \sqrt{-g} \left[ \frac{F(\sigma)}{2}R 
	- \frac{g^{\mu\nu}}{2} \partial_\mu \sigma \partial_\nu \sigma -\Lambda- V(\sigma) + {\cal L}_m \right] \,,
\end{equation}
where ${\cal L}_m$ is the matter Lagrangian.
Before considering any specific dynamics, let us note that the Planck mass $M_{\rm pl}^2 \equiv1/(8\pi G)$, is replaced by a functional of the scalar field $\sigma$. Being a dynamical field, the Newton constant evolves with time in these scalar-tensor models. Specifically, we must distinguish between a cosmological Newton constant, defined as
 $G_N = 1/(8\pi F)$ and an {\rm effective} Newton constant given by\footnote{We note that the $G_{\rm eff}$ defined in Eq.~\eqref{eq:Geff} can in principle receive space-dependent corrections when the scalar field is massive, see e.g.~\cite{Koyama:2015vza,Alby:2017dzl}. However, this is not the case in the model considered here.} \cite{Boisseau:2000pr}:
\begin{equation}
	\label{eq:Geff}
	G_{\rm eff} = \frac{1}{8\pi F}\frac{2F+4F^2_{\sigma}}{2F+3F^2_{\sigma}} \,.
\end{equation}
While $G_N$ is actually relevant for our study, as it is the one affecting cosmological observables, $G_{\rm eff}$ is constrained by laboratory experiments to be close to $1$ (in $M_{\rm pl}=1$ units, which we will adopt throughout this paper), with a precision of $1$ part in $10^5$. In addition, other experimental constraints on the theory come from bounds on the so-called Post-Newtonian (PN) parameters, used to phenomenologically capture  deviations from general relativity (GR) \cite{Will:2014kxa}. For the Lagrangian in  Eq.~\eqref{eq:model}, they are given by  \cite{Boisseau:2000pr}: 
\begin{align}
	\label{eqn:gammaPN}
	\gamma_{\rm PN}&=1-\frac{F_{\sigma}^{2}}{F+2F_{\sigma}^{2}},\\
	\label{eqn:betaPN}
	\beta_{\rm PN}&=1+\frac{F F_{\sigma}}{8F+12F_{\sigma}^{2}}\frac{\dd\gamma_{\rm PN}}{\dd\sigma},
\end{align}
and are constrained $\beta_{\rm PN}-1 = (4.1\pm7.8) \times 10^{-5}$ at 68\% confidence limits (C.L.) 
\cite{Will:2014kxa} and 
$\gamma_{\rm PN}-1 = (2.1 \pm 2.3) \times 10^{-5}$ at 68\% C.L. \cite{Bertotti:2003rm}. As, we will comment shortly, both PN and laboratory constraints are automatically satisfied by the dynamics we consider in this paper. 

The Klein-Gordon (KG) and Friedmann equations derived from the Lagrangian are respectively \footnote{We use over dots to denote derivatives with respect to cosmic time.}
\ba
\label{eqn:KGFR}
&\ddot{\sigma}+3H\dot{\sigma}+V_{\sigma} = \frac{1}{2}F_{\sigma}R ,\\
\label{eqn:NMC_Friedmann}
&3H^{2}F = \rho+\frac{\dot{\sigma}^{2}}{2}+V(\sigma)-3H\dot{F} \equiv\rho+\rho_{\rm scf}.
\ea
Our EMG model corresponds to the simple choice  $F(\sigma)=M_{\rm pl}^2+\xi\sigma^2$ for the NMC and  $V(\sigma)=\lambda \,\sigma^4/4$ for the potential, where both $\xi$ and $\lambda$ are dimensionless constants. For convenience, we define $V_0$ through $\lambda = 10^{2V_0}/(3.516\times 10^{109})$ as in Ref.~\cite{Braglia:2020auw}.

\begin{figure}
    \centering
    \includegraphics[width=\columnwidth]{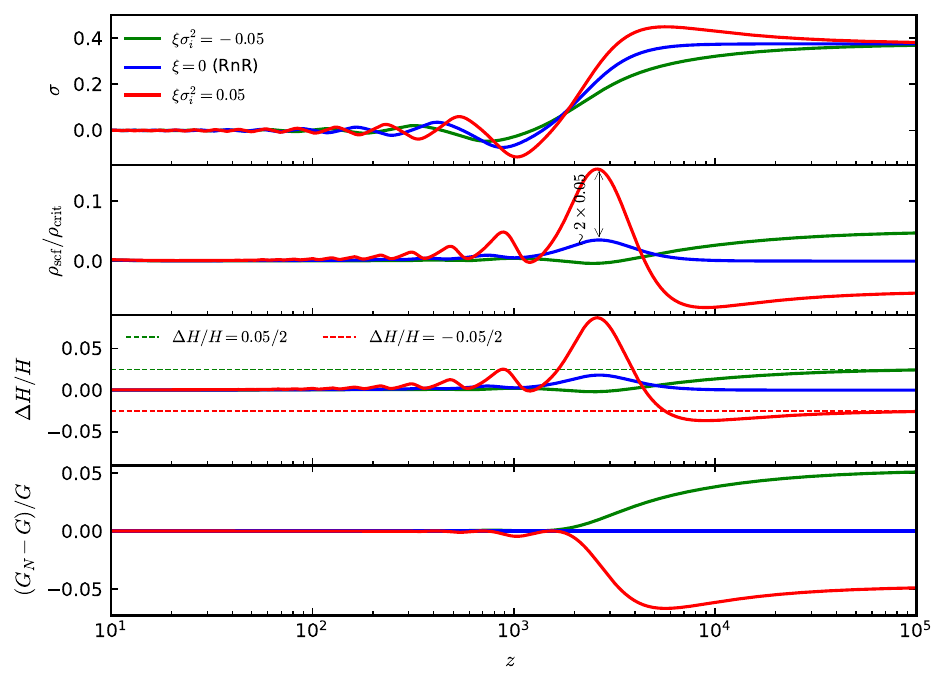}
    \caption{From the top to the bottom: evolution of the scalar field $\sigma$, fractional energy density of the scalar field $\rho_{\rm scf}/\rho_{\rm crit}$, fractional change in the Hubble rate $\Delta H/H$ with respect to $\Lambda$CDM for the same choice of cosmological parameters, and shift in cosmological Newton constant $(G_N-G)/G$. We consider three values of the NMC: $\xi\sigma_i^2 = -0.05$ (green), $\xi = 0$ (blue), and $\xi\sigma_i^2 = 0.05$ (red). We use the same value of $\sigma_i=0.375$ and $V_0=2$ for all the cases. In the third panel, the horizontal lines indicate the values of $\Delta H/H$ at $z \gg z_{\rm eq}$ (for the EMG model).}
    \label{fig:intro}
\end{figure}

The cosmological evolution of $\sigma(z)$ in our model is shown in Fig.~\ref{fig:intro} and goes as follows. At very early times, prior to the redshift of matter-radiation equality $z_{\rm eq}$, the scalar field $\sigma$ is frozen to its initial condition $\sigma_i$ due to Hubble friction. At such early times, the term $3 H \dot{F}$ in the Friedmann equation is negligible, and the effect of the frozen scalar field is two-fold. On one hand, it contributes to a shift in the cosmological Newton constant as $G_N(z\gg z_{\rm eq})/G_N(z=0)-1\sim-\xi\sigma_i^2$. On the other, it suppresses the Hubble rate, compared to the standard $\Lambda$CDM evolution as:
\begin{equation}
\label{eq:Hubble}
	\frac{H^{\rm EMG}-H^{\rm \Lambda CDM}}{H^{\rm \Lambda CDM}}(z\gg z_{\rm eq})\sim-\frac{\xi\sigma_i^2}{2},
\end{equation} 
as can be seen from the third panel in Fig.~\ref{fig:intro}.
Note that these estimates are derived under the assumption $\xi\sigma_i^2\ll1$, as we will show that these are the typical values of the model parameters favored by the data.
One can thus see that the combination $\xi\sigma_i^2$ enters in both the fractional variation of the cosmological Newton constant, as well as the one of the Hubble rate, hence being a genuine effective parameter that can be constrained by cosmological data. This property was indeed used to constrain the time evolution of $G_N$ from Big Bang nucleosynthesis (BBN) in several papers - see e.g. Ref.~\cite{Alvey:2019ctk} for the most updated constraints - and in previous CMB studies \cite{Umilta:2015cta,Rossi:2019lgt,Ballardini:2020iws,Braglia:2020iik,Ballardini:2021evv}. Furthermore, Eq.~\eqref{eq:Hubble} already explains why this class of models was proposed as explanation for $H_0$
tension~\cite{Rossi:2019lgt,Ballesteros:2020sik,Braglia:2020iik,Adi:2020qqf}. Depending on the sign of $\xi$, this EMG model can indeed decrease ($\xi>0$) or increase ($\xi<0$) the Hubble rate at early times with a corresponding increase/decrease in $r_s$ respectively.  Note that for $\xi > 0$, 
the scalar field acts as a negative energy component. While this may seem counterintuitive, we note that there is nothing wrong with it, as the energy density defined above is just an effective quantity.

The scalar field has an effective mass squared given by $m_{\sigma,\,{\rm eff}}^2\simeq V_{\sigma\sigma}-\xi R$.  As we get closer to $z_{\rm eq}$, the source term $-\xi R$ on the right hand side of the KG equation is slowly turned on ($R$ is identically zero for a Universe filled only with radiation) and drives the scalar field to larger/smaller values for $\xi>0$ and $\xi<0$ respectively. At some critical redshift $z_c$, when the effective mass of the scalar field becomes dominated by the $V_{\sigma\sigma}$ term and starts to be comparable to the Hubble rate, $\sigma$ falls down the potential and undergoes damped oscillations around the minimum at $\sigma=0$. The velocity of the scalar field increases quickly, leading to a sharp energy injection into the cosmic fluid, resembling that occurring in EDE models~\cite{Poulin:2018dzj}. Compared to minimally coupled EDE models, however, the maximum scalar field energy fraction $f_{\rm scf}\equiv\rho_{\rm scf}(z_c)/\rho_{\rm crit}(z_c)$ is also affected by the NMC, and can be probed to be larger/smaller depending on whether $\xi$ is positive or negative respectively. In particular, in the positive NMC case, a simple relation $f_{\rm scf}\sim f_{\rm scf}(\xi=0)+2\,\xi\sigma_i^2$, can be derived (see the second panel of Fig.~\ref{fig:intro}). In the negative case, the energy injection is broader, and the profile of the EMG energy constrast resembles the one of an extra-relativistic component~\cite{Braglia:2020iik}.

As touched upon above, this dynamics provides a natural way to evade the late time PN and laboratory constraints. Indeed, the large quartic potential drives the scalar field to its minimum at $\sigma\sim0$, where late time modifications to gravity are so small that they are well within the error bars from the constraints mentioned above.

Finally, let us mention that other regions of the parameter space are also possible. For example, the value of the effective mass squared $V_{\sigma\sigma}$ could
be large enough so that the scalar field rolls down the potential very early. In that case, $\sigma( z\gg z_{\rm eq}) \sim 0$ and $F_\sigma( z\gg z_{\rm eq}) \sim 0$. This situation is closer to $\Lambda$CDM than the current setting.
Another possibility is that the effective mass is so small that it is always subdominant with respect to $F_\sigma \,R$, as the limiting case of Eq. \ref{eq:model} with
$V=0$ in Ref.~\cite{Braglia:2020auw} or as in Refs.~\cite{Rossi:2019lgt,Ballardini:2016cvy,Ballardini:2020iws, Ballardini:2021evv} with different theoretical priors.
In this paper, motivated by results from past literature attempting at resolving the Hubble
tension \cite{Braglia:2020auw}, we will only consider
positive values of the coupling $\xi$.

\subsection{Analysis setup}
\label{sec:data}
Our goal is to perform an in depth study of the impact of the new small-scale measurements by ACT and SPT on the EMG model, and its ability to alleviate the Hubble tension. We therefore will consider various combinations of the following datasets:

\begin{itemize}
   \item {\it Planck:} The low-$\ell$ TT, EE and  the high-$\ell$ TT, TE, EE power spectra of the CMB from {\em Planck} DR3 \cite{Planck:2019nip}, covering $\ell\in[2:2500]$ in temperature and $\ell\in[2:2000]$ in polarization. When specified, we also include the `conservative' lensing potential reconstructed  from {\em Planck} DR3 \cite{Planck:2018lbu}.
    \item {\bf ACT:}  The TT, TE, EE power spectra of the CMB from ACT DR4 \cite{ACT:2020frw}, covering the range  $\ell\in[500:4000]$ in temperature 
    and $\ell\in[350:4000]$ in polarization. When combining with {\em Planck} data, we remove the TT data below $\ell = 1800$ as per recommendation from the ACT collaboration to avoid correlations between the two data sets. 
  \item {\bf SPT:} The TT, TE, EE  power spectra of the CMB measured with SPT-3G 2018 \cite{SPT-3G:2022hvq}, covering the range $\ell\in[750:3000]$ in temperature and $\ell\in[300:3000]$ in polarization. In this case, we do not remove any multipole when combining with other CMB experiments, following again the SPT collaboration recommendation.
    \item {\bf BAO/$f\sigma_8$:} The measurements of the BAO and redshift space distortion `$f\sigma_8$' from the CMASS and LOWZ galaxy samples of BOSS DR12 at $z = 0.38$, 0.51, and 0.61 \cite{BOSS:2016wmc} and the BAO measurements from 6dFGS at $z = 0.106$ and SDSS DR7 at $z = 0.15$  \cite{Beutler:2011hx,Ross:2014qpa}.

    \item {\bf Pan+:}  The Pantheon+ SN Ia catalog that compiles information about the luminosity distance to over 1550 SN Ia  in the redshift range  $0.001 < z < 2.3$ \cite{Brout:2022vxf}. 
    This is often also labelled as PanPlus.
    
     \item {\bf S$H_0$ES:} The S$H_0$ES determination of the SN Ia magnitude $M_b=-19.253 \pm
0.027$ \cite{Riess:2021jrx} modeled as a Gaussian likelihood. This 
 is simply referred as $M_b$ in the tables' caption.

\end{itemize}

We perform MCMC explorations of the parameter space with the code \texttt{MontePython-v3} \cite{Audren:2012wb,Brinckmann:2018cvx}, interfaced with a version of the \texttt{hi\_class} code \cite{Lesgourgues:2011re,Blas:2011rf,Zumalacarregui:2016pph,Bellini:2019syt} modified according to Ref.~\cite{Braglia:2020auw}. For all runs, we impose uninformative flat priors on $\{\omega_b,\,\omega_{\rm cdm},\,H_0,\,n_s,\,\ln{(10^{10} A_s)},\,\tau_{\rm reio}\}$, which correspond, respectively, to the dimensionless baryon energy density, the dimensionless cold dark matter energy density, the Hubble parameter today, the spectral index and amplitude of the primordial power spectrum of scalar fluctuations (defined at the pivot scale $k_0 = 0.05 \ \mathrm{Mpc}^{-1}$), and the reionization optical depth. In addition, to describe the EMG model, we impose flat priors on $\sigma_i\in[0.03,0.9]$, $V_0 \in [0.6,3.5]$ and the combination $\xi\sigma_i^2 \in[0,0.2]$. As explained in Sec.~\ref{sec:model}, this combination of parameter controls both the amount of energy injection tied to the NMC, and the early-time modification to the cosmological Newton constant $G_N$, which is better constrained by the data than $\xi$. To avoid running with arbitrarily large $\xi$ when $\sigma_i$ is small, that can lead the code to numerical instabilities, we reject any point with $\xi \geq 2$, which would anyway produce an uncontrolled growth of $\sigma$ leading to an extremely bad fit to data, as mentioned in the previous section.

We follow the \textit{Planck} conventions for the treatment of neutrinos, including two massless and one massive species with $m_{\nu} = 0.06$ eV \cite{Planck:2018vyg}, and enforcing $N_{\rm eff}=3.046$. We include non-linear correction using the \texttt{halofit} prescription \cite{Smith:2002dz}.
Given the complexity of the parameter space in some analyses, when checking the convergence of the chains, we adopt a slightly looser  Gelman-Rubin \cite{Gelman:1992zz} convergence criterion, namely $R-1 < 0.1$, although most chains are converged well below the conventional $R-1 <0.05$.
Finally, we extract the best-fit parameters from the simulated annealing procedure described in details in the appendix D of Ref.~\cite{Schoneberg:2021qvd}, and estimate the preference for the extended model $\mathcal{M}$ (EMG or RnR) relative to $\Lambda$CDM using the Akaike Information Criterium (AIC):
\begin{equation}
\Delta {\mathrm{AIC}} = \chi^2_{\mathrm{min}} (\mathcal{M}) -\chi^2_{\mathrm{min}}(\Lambda{\rm CDM})+2(N_{\mathcal{M}} - N_{\Lambda \mathrm{CDM} } )
\label{eq:AIC}
\end{equation}
where $N_{\mathcal{M}}$ stands for the number of free parameters in each model. 

\section{Looking for EMG in cosmological data}
\label{sec:results}
\subsection{Hints of EMG in CMB data}
\label{sec:EMG_CMB}
\begin{figure}
    \centering
    \includegraphics[width=\columnwidth]{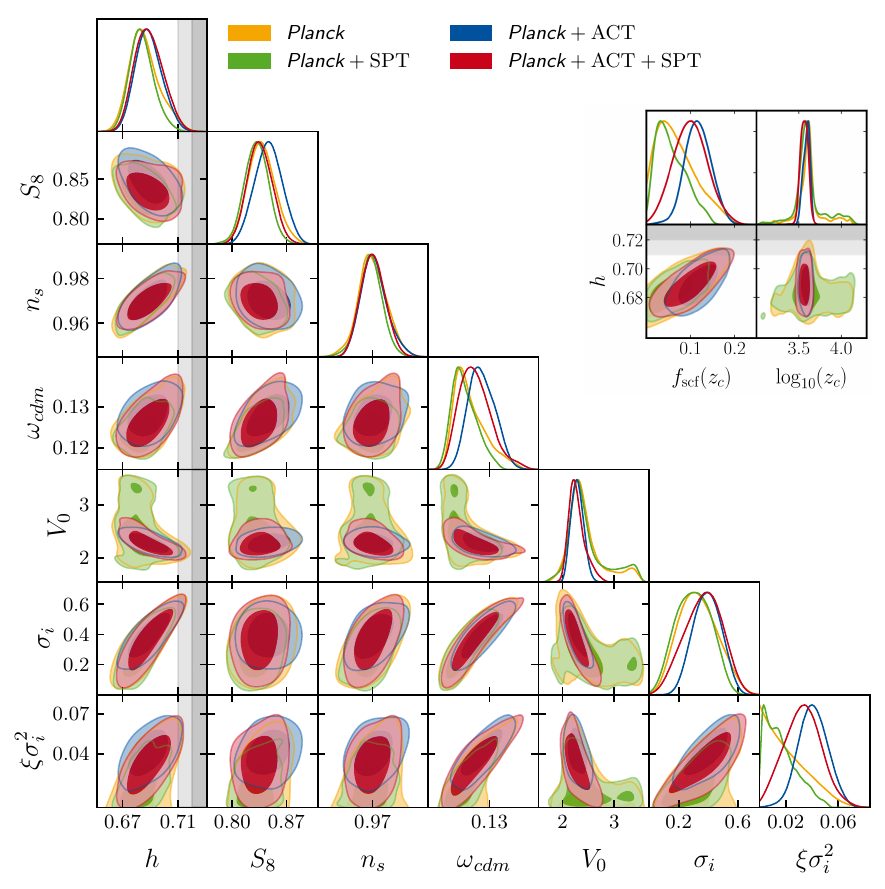}
    \caption{1D and 2D posterior distributions (68\% and 95\% C.L.) for a subset of the cosmological parameters for different
    combinations of CMB experiments fit to EMG. The vertical gray band represents the $H_0$ value $H_0 = 73.04 \pm 1.04\,{\rm km}\,{\rm s}^{-1}\,{\rm Mpc}^{-1}$ as reported by the S$H_0$ES collaboration \cite{Riess:2021jrx}. }
    \label{fig:EMG_Planck_ACT_SPTfull}
\end{figure}

In this section, we start by performing an analysis with {\em Planck} data alone, to compare with results from the literature and set a baseline for the addition of the small-scale CMB measurements. We then add small scale CMB experiments to the full {\em Planck} data, first independently and then together.  The results of our analyses are given in Tab.~\ref{tab:CMB} and displayed in Fig.~\ref{fig:EMG_Planck_ACT_SPTfull}. We also report constraints on $f_{\rm scf}(z_c)$ and $\log_{10}(z_c)$, computed as derived parameters (note that $z_c$ is defined as the redshift at which the scalar field energy fraction is maximum). \\

{\bf Results with {\em Planck}:} Starting with {\it Planck} alone, we find a marginally significant detection of $f_{\rm scf}(z_c)=0.070^{+0.028}_{-0.055}$ at $\log_{10}(z_c) = 3.626^{+0.053}_{-0.12}$  at 68\% C.L., with a small improvement in {\em Planck} $\chi^2$, with $\Delta \chi^2 \simeq -4$ compared to $\Lambda$CDM. However, we only derive an upper limit on $\xi\sigma_i^2 < 0.054$ (95\% C.L.). A quick estimate of the significance of the detection via the $\Delta {\mathrm{AIC}} \simeq +2$ shows that the model is weakly disfavored over $\Lambda$CDM due to the three additional parameters.
Note that these bounds differ slightly from what was reported in Ref.~\cite{Schoneberg:2021qvd}, that considered a similar data combination. This is due to our choice of considering  $\xi\sigma_i^2$ as a free parameter in the run (rather than $\xi$), motivated by our discussion in Sec.~\ref{sec:model}. This allows to probe more effectively the relevant phenomenological parameter combination. However, we also note that this creates a different prior volume in $\sigma_i$, since too small values of $\sigma_i$ may lead to large values of $\xi$ that are excluded by the data. Therefore, we do not read into the apparent detection of $\sigma_i$ reported in Tab.~\ref{tab:CMB}, that is not supported by the $\Delta\chi^2$. On the other hand, (the lack of) constraints to $V_0$ appear in reasonable agreement with the results of Ref.~\cite{Schoneberg:2021qvd}. \\

{\bf Results with {\em Planck}+SPT:} When SPT data are added to the analysis, constraints become slightly tighter, although the mild detection is still present as induced by the (small) improvement in the fit to {\em Planck} data. We find  $f_{\rm scf}(z_c)=0.056^{+0.022}_{-0.044}$ at $\log_{10}(z_c) = 3.62^{+0.06}_{-0.15}$  (68\% C.L.),
and a tighter bound on  $\xi\sigma_i^2 < 0.04$ (95\% C.L.). Similarly, constraints to $\sigma_i$ and $V_0$ are only mildly affected by the inclusions of SPT.\\

{\bf Results with {\em Planck}+ACT:} 
When ACT is added to {\em Planck}, the situation is quite different. In that case, one can see that there is a clear detection of the EMG  model, with $f_{\rm scf}(z_c) =0.117\pm 0.030$ around  $\log_{10}(z_c) = 3.589^{+0.046}_{-0.034}$ and a detection of the nonminimal coupling $\xi\sigma_i^2 =0.041\pm 0.011$. 
The $\chi^2$ of the data is improved by $\Delta\chi^2 = -10.24$, while the model is weakly favored over $\Lambda$CDM according to $\Delta\mathrm{AIC}=-4.24$.\\

{\bf Results with {\em Planck}+SPT+ACT:} 
When considering {\em Planck}+SPT+ACT, we still find a detection of the EMG model with $f_{\rm scf}(z_c) =0.100\pm 0.039$, $\log_{10}(z_c) = 3.564\pm 0.045$ and  $\xi\sigma_i^2 = 0.033\pm 0.014$. Although the significance is slightly reduced from the analysis of {\em Planck}+ACT, we still find an improvement in the fit over $\Lambda$CDM with  $\Delta\chi^2 = -8.12$ and $\Delta\mathrm{AIC}=-2.12$. \\

Comparing with results from the literature, the detection of an EMG contribution resembles the results found in the axion-like EDE model \cite{Hill:2021yec,Poulin:2021bjr,Smith:2022hwi}. However, in the case of the axion-like EDE model, the contribution of EDE is detected only when {\em Planck} temperature high-$\ell$ data
are cut at multipole $\ell > 1300$. When these data are included, the EDE contribution is found to be compatible with zero at $1\sigma$ in a Bayesian analysis. 
Here, despite the inclusion of the full dataset from {\em Planck}, and the full SPT temperature data\footnote{We note that current available analysis of the axion-like EDE model do not use SPT temperature data, but work in progress \cite{Smith:InPrep} indicate that the inclusion these data only strengthen constraints on EDE.}, one finds that a non-zero EMG contribution is favored at the $\gtrsim 2\sigma$ level. In the next section, we investigate why the NMC plays a strong role in favoring the EMG model over a rolling scalar field EDE model.

\begin{figure}[t!]
    \centering
    \includegraphics[width=\columnwidth]{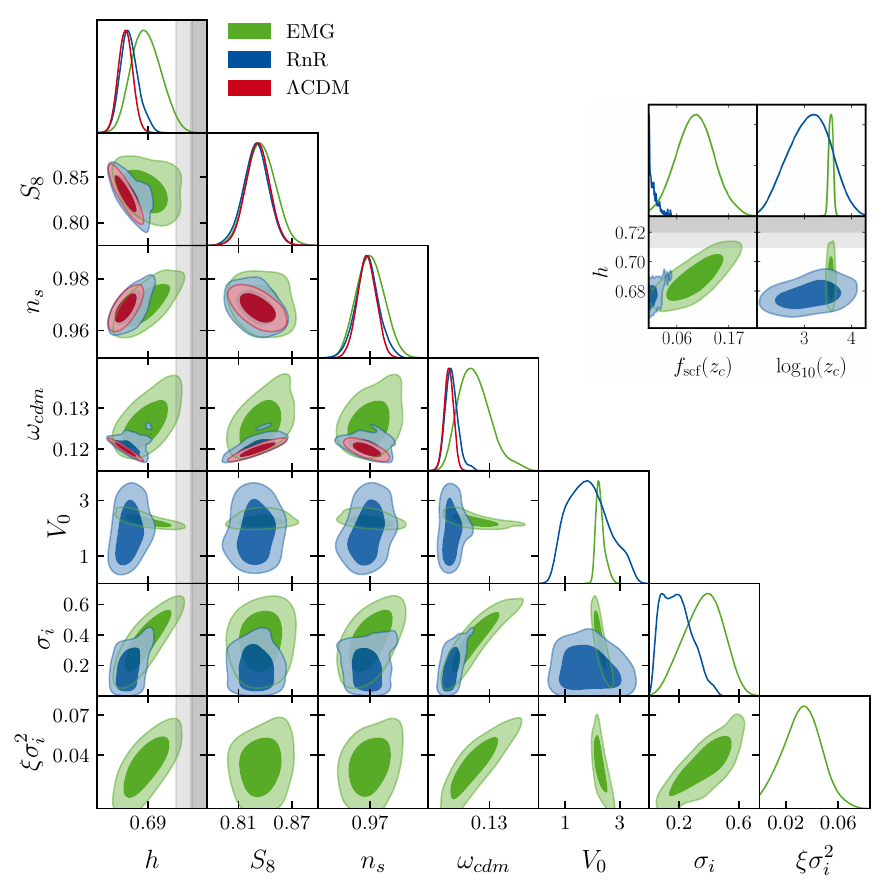}
    \caption{ 1D and 2D posterior distributions (68\% and 95\% C.L.) for a subset of the cosmological parameters for the dataset \emph{Planck}+ACT+SPT
    fit to three different models: $\Lambda$CDM, RnR, and EMG. The vertical gray band represents the $H_0$ value $H_0 = 73.04 \pm 1.04\,{\rm km}\,{\rm s}^{-1}\,{\rm Mpc}^{-1}$ as reported by the S$H_0$ES collaboration \cite{Riess:2021jrx}.  }
    \label{fig:EMG_vs_RnR}
\end{figure}


\subsection{Comparison with the Rock'n'Roll EDE model}
\label{sec:EMG_RnR}

\begin{figure}
    \centering
       \includegraphics[width=0.86\columnwidth]{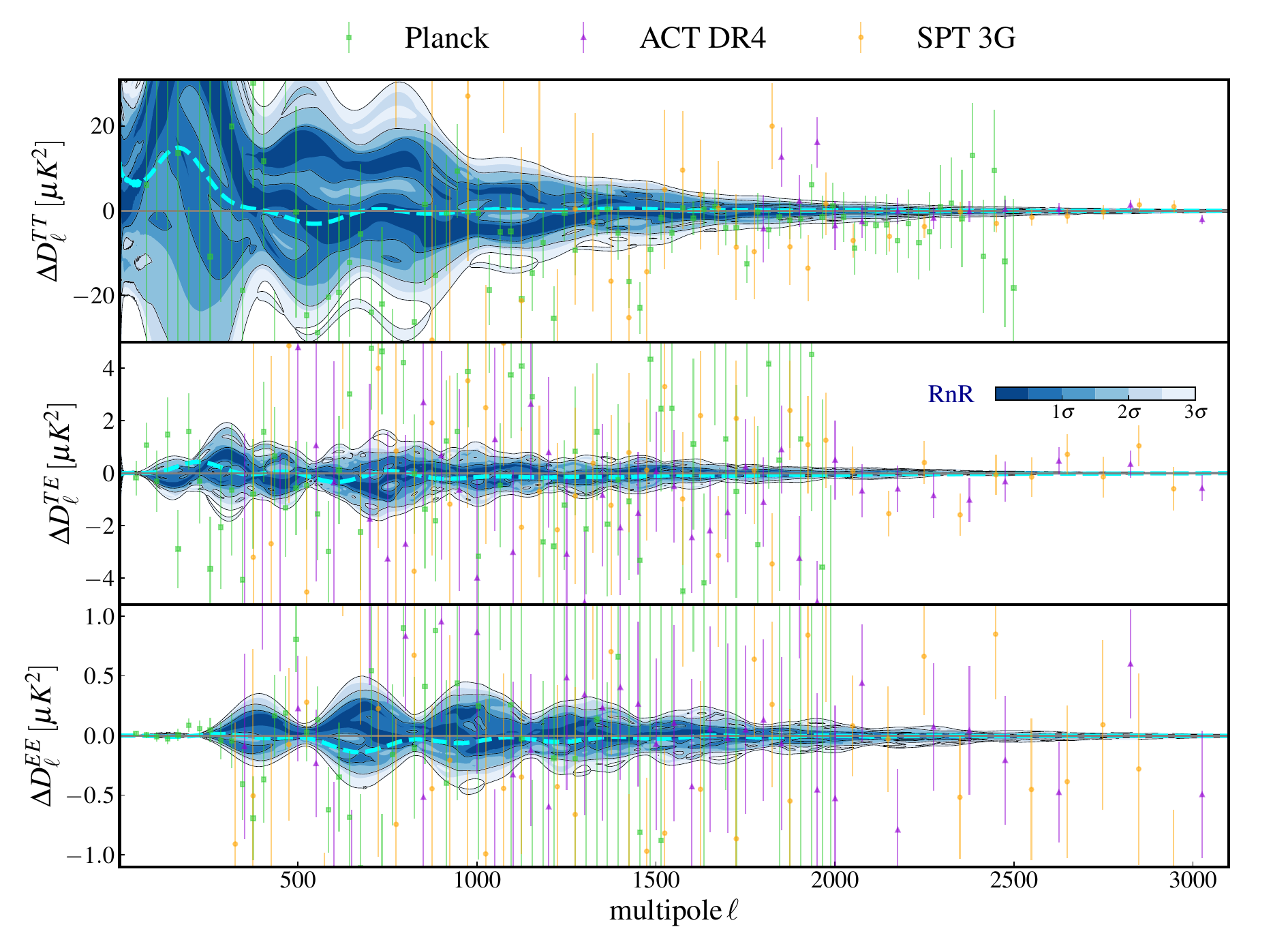}
        \includegraphics[width=0.86\columnwidth]{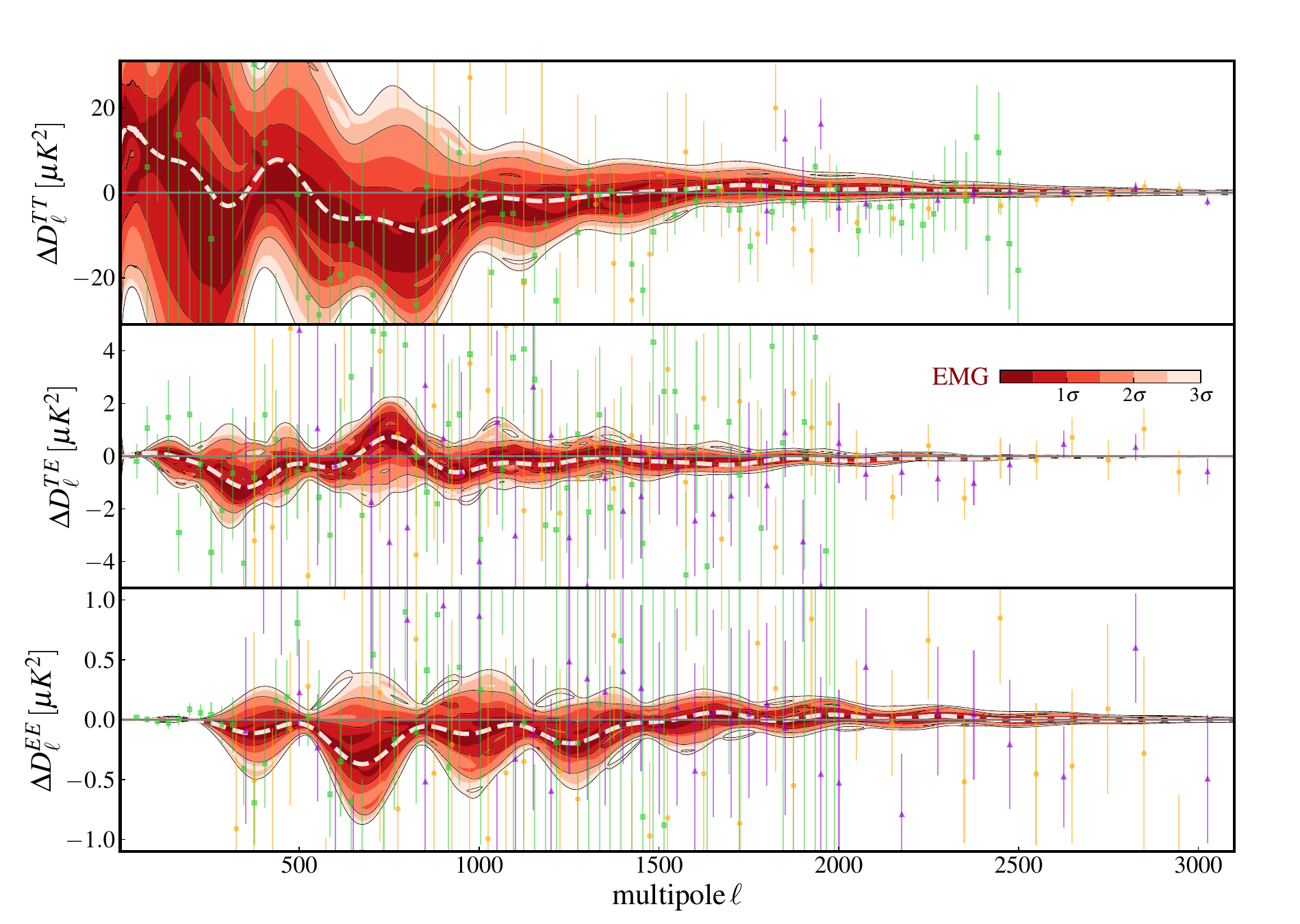}
    \caption{Functional posterior for the CMB residual spectra in the RnR (upper) and EMG (lower) models fit to the data combination \emph{Planck}+ACT+SPT. We additionally show the data residuals (coloured data points) and the best-fit of RnR (dashed blue) and EMG (dashed pink). All the residuals are computed with respect to the $\Lambda$CDM best-fit of the same data combination. }
    \label{fig:residuals_emg}
\end{figure}

In order to highlight the importance of the NMC in the preference for EMG, as opposed to a simple EDE contribution, we perform an analysis of {\em Planck}+ACT+SPT data in which we set $\xi=0$. As discussed previously, the model with $\xi=0$ matches the Rock'n'Roll EDE model, and we refer to it as RnR.
The results are presented in Fig.~\ref{fig:EMG_vs_RnR} and Tab.~\ref{tab:CMB_2}. 
One can see that the data shows no preference for the RnR model, and in particular the fractional contribution of EDE is constrained  $f_{\rm scf}(z_c) < 0.034$. This is significantly different than what we reported in the EMG case, with $f_{\rm scf}(z_c) =0.0995\pm 0.039$, and a clear detection of $\xi\sigma_i^2 = 0.033\pm 0.014$.
This raises the question: {\it why does the NMC allow to improve the fit to the data significantly over the RnR model?} 
To get a first hint at the answer to this question, we show the CMB residuals (the band represents the functional posteriors and the dashed line the best-fit) of the EMG and RnR models fit to {\em Planck}+SPT+ACT in Fig.~\ref{fig:residuals_emg}. 
While the models provide similar residuals $\ell\gtrsim 1500$, and stay compatible with zero at $\gtrsim2\sigma$ at all multipoles,  one can see that the EMG model has a significant dip in TT and EE in the region $\ell \sim 500-1000$ \footnote{Looking at Fig.~\ref{fig:residuals_emg}, and similarly to the discussion in Ref.~\cite{Smith:2022hwi}, we see that the dip in TT at $\ell \sim 750$ is driven by the amplitude mismatch between \emph{Planck} and ACT DR4 at high-$\ell$, which has a larger amplitude. SPT does not show this mismatch with \emph{Planck}, which explains why the preference for EMG is only seen in the analyses including \emph{Planck}+ACT.}, that is not present in the RnR model. We now investigate the reason for that difference.

First, let us stress that the EMG best-fit has a larger $H_0$ compared to the RnR one, and a significantly larger $f_{\rm scf}$, such that some of the obvious features like the dip seen around $\ell\sim 750$ in TT should not necessarily be interpreted as a unique signature of EMG. Rather, the question is why can EMG attain such a larger contribution and $H_0$ value (together with other parameters reshuffling)  leading to an apparent dip, that cannot be attained in RnR?

\begin{figure}[t!]
	\centering
	\includegraphics[width=0.95\columnwidth]{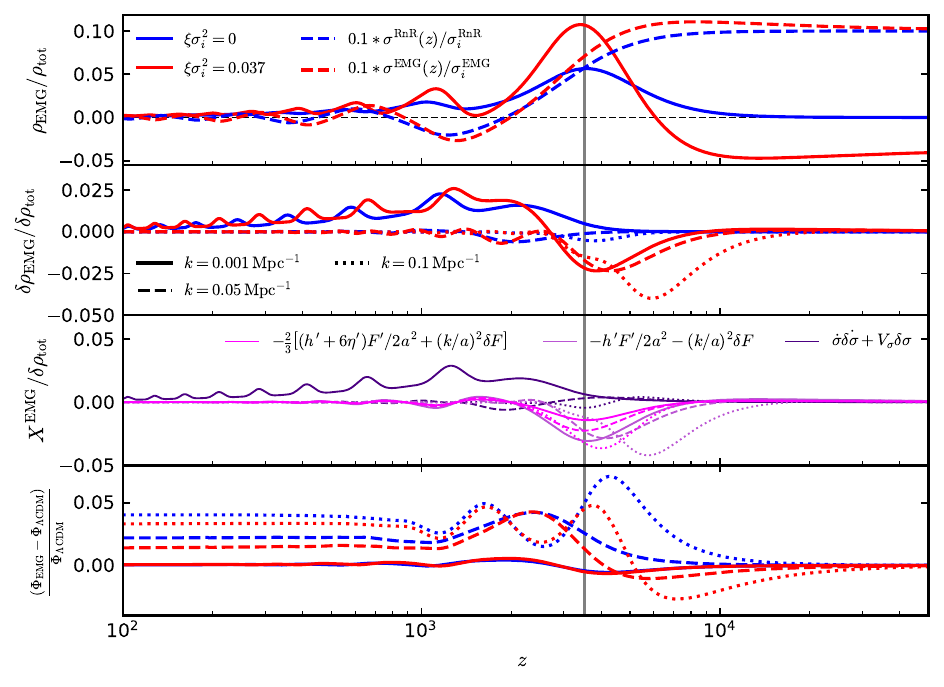}
	\caption{In this figure we adopt the EMG best-fit parameters from the {\em Planck}+ACT+SPT
        analysis. For RnR, we set the parameter $\xi=0$, and adjust $\sigma_i$ and $V_0$ in order to keep $r_s$ and $z_c$ to the EMG best-fit values. This ensures that the main background effects are similar in both models.  [$1^{\rm st}$ panel.] Evolution of the energy density fraction and the scalar field. [$2^{\rm nd}$ panel.] Evolution of the density perturbations in the synchronous gauge for 3 modes. [$3^{\rm rd}$ panel.] We show the quantities in the legend normalized to $\delta\rho_{\rm tot}$ in the EMG case (see main text for definitions). [$4^{\rm th}$ panel.] Fractional variation in the evolution of the Weyl potential with respect to the $\Lambda$CDM best-fit to \textit{Planck}+SPT+ACT.}
	\label{fig:perturbations}
\end{figure}

In order to answer the question, we need to find a way to isolate the effect of the NMC at the level of the CMB residuals. This task is not trivial, as the model is described by 3 free parameters ($\sigma_i,\,V_0$ and $\xi \sigma_i^2$), and several combinations of them (together with the other 6 cosmological parameters) can conspire to create a given feature in the residuals. However, we can get some intuition about the role of the NMC by proceeding as follows. Starting from the EMG best-fit model to \textit{Planck}+SPT+ACT, we set $\xi\sigma_i^2=0$, and adjust $V_0$ and $\sigma_i$ (all other parameters are kept unchanged) so that we recover the same value of $r_s = \int^{\infty}_{z_{\rm dec}}\dd z\, c_s(z)/H(z)$. We now are in a situation where both models lead to the same $H_0$, and the same angular size of the sound horizon $\theta_s$, thereby removing the dominant effects induced by differences at the background level. The corresponding dynamics in the EMG and RnR are shown in the top panel of Fig.~\ref{fig:perturbations}, in blue and red color respectively. 
 As can be seen, the NMC leads to a sharper and stronger energy injection around $z_c$ compared to its RnR counterpart. However, the effective energy density of the scalar field is negative at early times, suppressing the Hubble rate at $z>z_c$. The two effects compensate in the integral to arrive at $r_s$ so that $r_s^{\rm EMG}\sim r_s^{\rm RnR}$. 
 Let us also note that, as explained in Ref.~\cite{Braglia:2020auw}, the positive NMC $\xi>0$ makes the scalar field grow at early times as we get closer to the matter-radiation equality (dashed black line). Because of that, when the effective mass of $\sigma$ becomes larger than $H$, the scalar field falls down the potential from a value $\sigma(z\gtrsim z_c)>\sigma_i$. 
 
The picture described above leads to a peculiar feature of the evolution of the density perturbations, shown in the central panel in Fig.~\ref{fig:perturbations}. There, we plot the evolution for 3 modes that cross the Hubble radius well before ($k=0.1$ Mpc$^{-1}$), shortly before\footnote{The mode entering the Hubble radius at $z_c$ is $k\sim0.01$ Mpc$^{-1}$. We choose to use $k=0.05$ Mpc$^{-1}$ in our analysis as it is the one corresponding to the $\ell\sim750$ dip in the TT spectrum.} ($k=0.05$ Mpc$^{-1}$) and long after ($k=0.001$ Mpc$^{-1}$) $z_c$, similarly to the analysis in Ref.~\cite{Poulin:2023lkg}. As we can see, the effective density perturbations in the  EMG sector feature a dip in their evolution at early times $z\gtrsim z_c$.  The subsequent evolution for $z\lesssim z_c$ is also slightly different from the minimally coupled case, but this is mostly due to the small differences in the background due to the different oscillatory phases. This can be understood by looking at the third panel in Fig.~\ref{fig:perturbations}: the dip in the evolution is driven by the terms proportional to $k^2 H\delta F$ and $F'\,h'$ in the EMG perturbations $\delta\rho_{\rm EMG}$ at $z>z_c$ (see Eq.~\eqref{eq:deltarhoemg}), and become quickly negligible for $z<z_c$ during the coherent oscillations of $\sigma$ around the minimum of the potential. The two terms affect all the modes, but due to the $k^2$ suppression, they are mostly important for the modes inside the horizon, like the dotted lines in the Figure. Note that the terms are also responsible for early time anisotropic stresses (magenta lines in Fig.~\ref{fig:perturbations}). The latter are slightly different from the extra term in the density perturbations due to the $6\eta'$ factor (see Eq.~\eqref{eq:anisotropicemg}).  As a result of the modified dynamics, the Weyl potentials (shown in the bottom panel of the first plot) are suppressed for the modes that are inside the horizon, compared to their RnR counterparts.

\begin{figure}[t!]
	\centering
	\includegraphics[width=0.95\columnwidth]{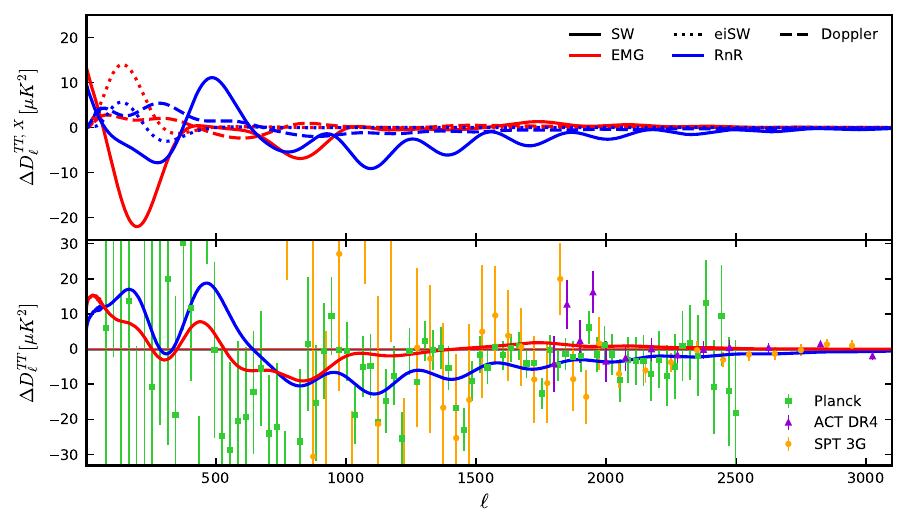}
	\caption{Difference of CMB TT spectra between the EMG/RnR and $\Lambda$CDM models. In this figure we adopt the EMG best-fit parameters from the {\em Planck}+ACT+SPT run. For RnR, we set the parameter $\xi=0$, and adjust $\sigma_i$ and $V_0$ in order to keep $r_s$ and $z_c$ to the EMG best-fit values.  For  $\Lambda$CDM we take the best-fit to {\em Planck}+ACT+SPT.  The data residuals (coloured data points) are computed with respect the $\Lambda$CDM best-fit of the same data combination . In the top panel we plot the {\em unlensed}  Sachs-Wolfe, early integrated Sachs-Wolfe (eISW), and Doppler contributions to the TT spectra, while the bottom panel shows the full TT spectra.}
	\label{fig:cmb}
\end{figure}

We now turn to describing the effect of the NMC on the CMB power spectra.  We show the difference of CMB power spectra in the EMG and RnR model with respect to the fiducial $\Lambda$CDM model in Fig.~\ref{fig:cmb}, in blue and red respectively. Note that we take as fiducial $\Lambda$CDM model the best-fit to \emph{Planck}+ACT+SPT, which has therefore different values for the six standard cosmological parameters than the EMG and RnR model.
As shown in the top panel, the main difference between EMG and RnR is a less-suppressed Sachs-Wolfe (SW) effect in the TT spectra at $\ell\gtrsim 1000$, and the absence of a bump visible around $\ell \sim 500$, that is driven by the dynamics of the Weyl potential described above \footnote{As we expand upon in App.~\ref{app:SW}, the pre-recombination time evolution of the Weyl potential is key in order to understand the impact of EMG and RnR on the SW effect, as the latter depends explicitly on integrated terms over a wide redshift range (see Fig.~\ref{fig:SW_approximate}).}. Similarly, the doppler term shows a net suppression in the RnR model, not seen in the EMG, while the eISW contribution is also slightly smaller around $\ell\sim100$ in RnR, as one can see that the Weyl potentials have a narrower time evolution in EMG than in RnR. 
The overall effect results in a suppressed power at $\ell>750$ in the RnR model, and a higher bump at $\ell \sim 200-500$. In fact, it is important to note that the RnR can indeed lead to a dip, but significanlty broader in multipoles than in the EMG model.  

In conclusions, the RnR model that would achieve the same $\theta_s$ and $H_0$ leads to a strongly suppressed Sachs-Wolfe term compared to the $\Lambda$CDM fiducial, that ruins the fit to multipoles $\ell \gtrsim 750$ in TT. For this reason, the best-fit for RnR is characterized by a small $f_{\rm scf}$ so that its effects on the CMB spectra are minimized. On the other hand,  in the EMG model, the NMC allows to decrease the impact on the driving of the CMB source terms (see App.~\ref{app:SW}), thereby reducing the impact on the Sachs-Wolfe effect. The net result is a dip around $\ell\sim750$ (that comes both from the reshuffling of $\Lambda$CDM parameters and the impact of EMG itself), keeping at the same time under control the residuals at larger multipoles, resulting in a preference for a nonminimal coupling.

\subsection{Impact of external datasets and tension with S$H_0$ES}
\label{sec:EMG_Ext}

\begin{figure}[t!]
    \centering
    \includegraphics[width=\columnwidth]{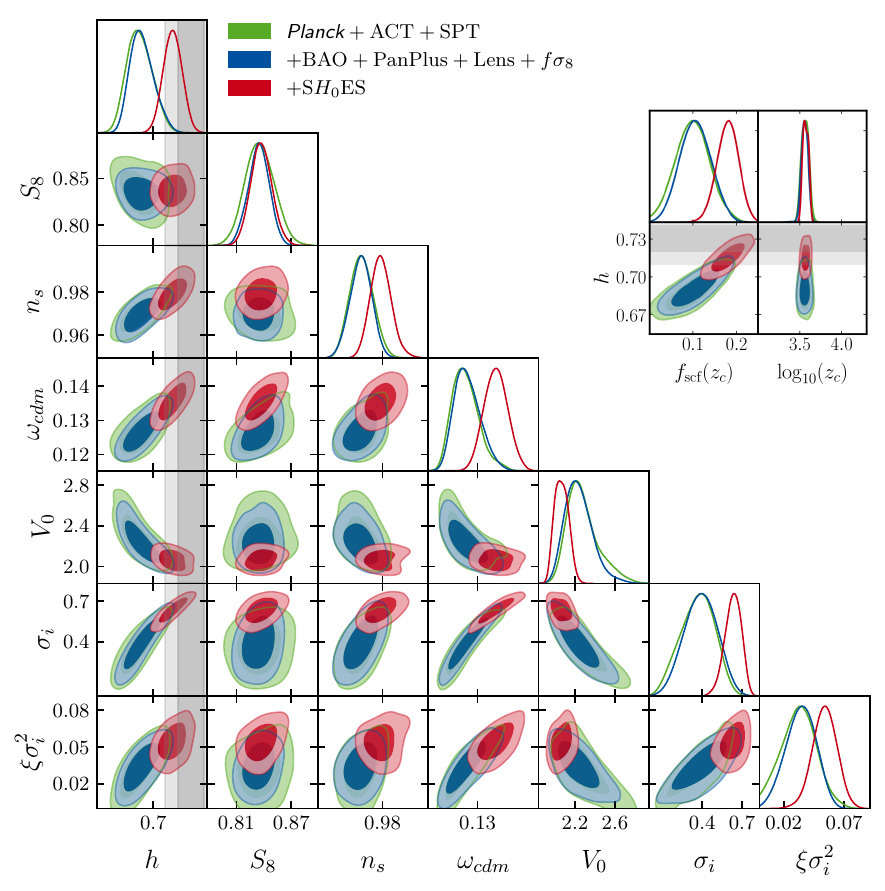}
    \caption{1D and 2D posterior distributions (68\% and 95\% C.L.) for a subset of the cosmological parameters for different data combinations fit to EMG. The vertical gray band represents the $H_0$ value $H_0 = 73.04 \pm 1.04\,{\rm km}\,{\rm s}^{-1}\,{\rm Mpc}^{-1}$ as reported by the S$H_0$ES collaboration \cite{Riess:2021jrx}. }
    \label{fig:EMG_ext}
\end{figure}
Until now, our discussion was limited to the study of the EMG model in light of CMB data. We now include external datasets listed in Sec.~\ref{sec:data} and compute the residual tension with S$H_0$ES. Results of analyses are shown in Fig.~\ref{fig:EMG_ext} and Tab.~\ref{tab:CMB_3}, and tension metrics (Gaussian tension and $Q_{\rm DMAP}$ as defined in Ref.~\cite{Schoneberg:2021qvd}) are provided in Tab.~\ref{tab:tension_metrics}. First, one can see that the preference for EMG is not affected by the inclusion of {\em Planck} DR3 lensing, BAO/$f\sigma_8$ and PanPlus data, with marginal changes in the reconstructed posteriors and $\chi^2$. Second, the inclusion of the S$H_0$ES $M_b$ prior drives the preference for EMG to above $5\sigma$, as expected given the tight correlation between $f_{\rm scf}$ and $H_0$. The $Q_{\rm DMAP}$ tension metric (more reliable than the Gaussian Tension given the non-gaussianity of the posterior \cite{Raveri:2018wln}) indicates that the ``Hubble tension'' reduces from $6.4\sigma$ \footnote{Note that our estimate of the tension in $\Lambda$CDM is larger than what is reported by the S$H_0$ES collaboration \cite{Riess:2021jrx}. This is due to the use of the full Pantheon+ dataset (that pulls $H_0$ and $\Omega_m$ up \cite{Brout:2022vxf}) and the inclusion of new CMB data.} in $\Lambda$CDM to $3.3\sigma$ in EMG, which while being a significant reduction, indicates that the EMG model cannot fully explain the tension. 
In addition, the value of the parameter $S_8\equiv \sigma_8 (\Omega_m/0.3)^{0.5}$, to which galaxy weak lensing surveys are particularly sensitive and that has been reported in $2-3\sigma$ tension with the $\Lambda$CDM prediction \footnote{We note that a recent combined analysis of KIDS and DES gives constraints on the $S_8$ parameter that are consistent with CMB data at the 1.7$\sigma$ level under $\Lambda$CDM \cite{Kilo-DegreeSurvey:2023gfr}. }
\cite{Heymans:2020gsg,DES:2021wwk,Dalal:2023olq}, is barely affected by the EMG model. This is in contrast with EDE that tends to slightly increase the $S_8$ tension \cite{Hill:2020osr,Murgia:2020ryi}. This effect was already observed in Ref.~\cite{Braglia:2020auw}, and can be attributed to the NMC that reduces the growth of perturbations and the amplitude of the Weyl potential, as we detailed above.

\section{Discussion and Conclusions}
\label{sec:concl}
The search for extensions to GR is among one of the most active field of research in modern days,  especially in the context of the dark matter and dark energy puzzle~\cite{Koyama:2015vza}. 
The recent tensions regarding measurements of some cosmological parameters \cite{Abdalla:2022yfr}, in particular the mismatch in the determination of the Hubble parameter between CMB (assuming $\Lambda$CDM) and cepheid-calibrated SN Ia \cite{Riess:2021jrx}, have raised the possibility that GR is indeed incomplete, as it has been shown that modifications to GR in the early universe are promising to resolve the Hubble tension \cite{Umilta:2015cta,Zumalacarregui:2016pph,Braglia:2020auw} (for discussion around modifying GR in the late-universe in the context of cosmic tensions, see Ref.~\cite{Pogosian:2021mcs}.)
In this paper, we have studied a realization of a scalar-tensor theory, dubbed early modified gravity or EMG model, where  a scalar field $\sigma$ with an effective mass induced by a quartic potential $V(\sigma) = \lambda \sigma^4/4$ is  non-minimally coupled to the Ricci curvature via $F(\sigma) = M_{\mathrm{pl}}^2+\xi\sigma^2$. 
Such a model was shown to successfully address the Hubble tension \cite{Braglia:2020auw,Schoneberg:2021qvd}, but new CMB data at small angular scales have been released, namely ACT DR4 and SPT-3G 2018 TT/TE/EE, that have the constraining power to test the EMG model. 
Until now, analyses of the EMG model in light of these data were lacking. 
Our main results are as follows:

\begin{itemize}
    \item When considering the combination of {\em Planck}+SPT+ACT data, we find a detection of the EMG model, characterized with a maximal fractional contribution to the energy density $f_{\rm scf}(z_c) =0.100\pm 0.039$ that is reached at  $\log_{10}(z_c) = 3.564\pm 0.045$, with a nonminimal coupling  $\xi\sigma_i^2 = 0.033\pm 0.014$, and a $\Delta\chi^2\simeq-8.1$ ($\Delta$AIC $\simeq-2.1$) compared to $\Lambda$CDM. 
    \item The preference for EMG is driven mostly by ACT data ($\Delta\chi^2\simeq-6$) and slightly by {\em Planck} high-$\ell$ ($\Delta\chi^2\simeq-3.5$). The $\chi^2$ of SPT is only marginally affected at the EMG best-fit.
    \item The detection of EMG is unnaffected by the inclusion of SN Ia, BAO/$f\sigma_8$ and {\em Planck} lensing data. We quantify a residual Hubble tension of $3.3\sigma$, strongly decreased from $6.4\sigma$ in $\Lambda$CDM (for the same data combination), but not fully alleviated.
  
\end{itemize}
Our results can be compared with popular `axion-like' EDE models, which show hint of a detection at $\sim 3\sigma$ when removing high-$\ell$ {\em Planck} temperature data, but  $f_{\rm EDE}(z_c)\sim0$ at $1\sigma$ with the full data \cite{Hill:2021yec,Poulin:2021bjr,Smith:2022hwi}. We find that the EMG model is fitting CMB data significantly better than the axion-like EDE model, and it would be interesting to extend our work to study a non-minimally coupled axion-like EDE model (or other form of the potential). It would also be interesting to explore whether the preference for EMG is affected by systematic errors, for instance following what was done for EDE, by changing the TE polarization efficiencies in \emph{Planck} data \cite{Smith:2022hwi} or rescaling the overall amplitude of ACT TE data \cite{Hill:2021yec}.

\begin{figure}[t!]
    \centering
    \includegraphics[width=0.49\columnwidth]{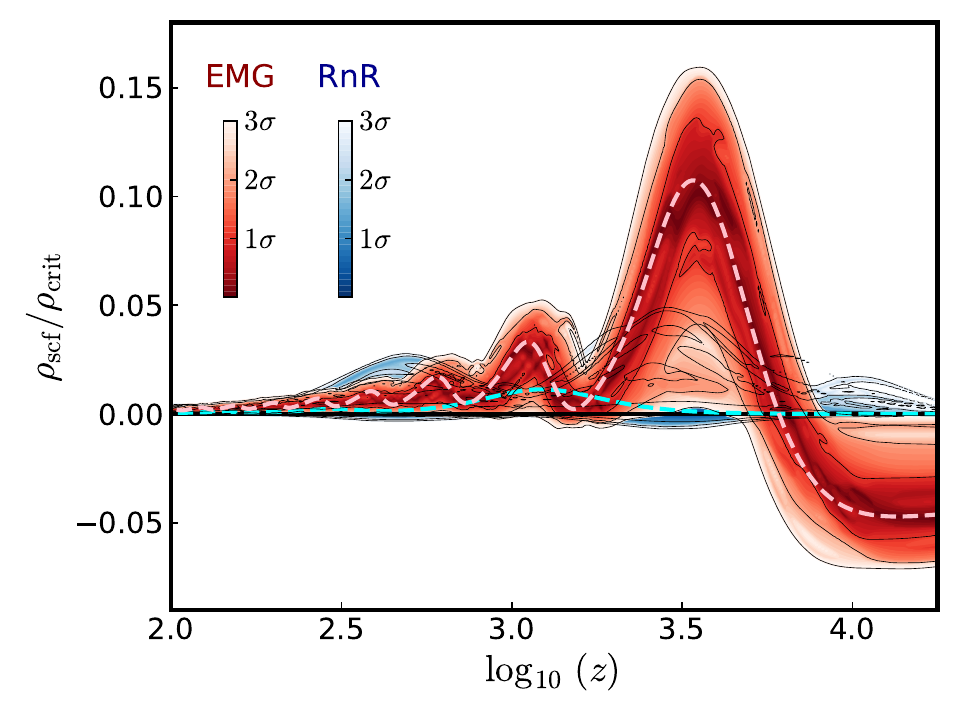}
        \includegraphics[width=0.49\columnwidth]{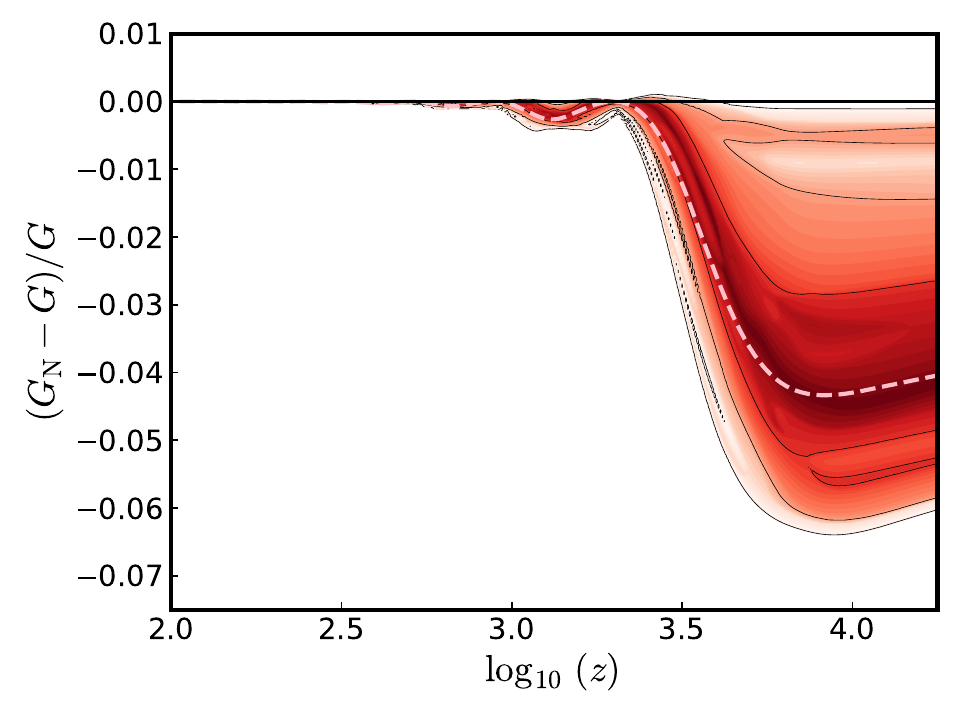}
    \caption{Functional posterior for the fractional energy density of the scalar field (in EMG and RnR) and for the shift in the cosmological Newton constant (in EMG) fit to the data combination \emph{Planck}+ACT+SPT. The dashed lines correspond to the best-fit. }
    \label{fig:predictive_posteriors}
\end{figure}

We have additionally investigated the role of the NMC in driving the preference for EMG. Indeed, when $\xi$ is set to $0$, the model is identical to the RnR EDE model. We show the fractional contribution of the scalar field to the total energy density as a function of redshift  in both EMG and RnR model in the left panel of Fig.~\ref{fig:predictive_posteriors}, as indicated by the combination of {\em Planck}+SPT+ACT. One can see that the RnR model is strongly constrained compared to EMG, and excluded as a resolution of the Hubble tension. 
To understand the reason for that preference for EMG over RnR, we have demonstrated that the RnR model that would achieve the same $\theta_s$ and $H_0$ strongly suppresses the Sachs-Wolfe term compared to best-fit $\Lambda$CDM model, and is excluded by TT data at multipole $\ell \gtrsim 750$ (mostly). In the EMG model, the inclusion of the NMC affects the time evolution of the Weyl potential in a way that yields a weaker driving of the CMB source terms, thereby reducing the impact on the Sachs-Wolfe term. 

Our results indicate that the cosmological Newton constant may have differed from  its canonical value in the past by up to $6\%$, as shown in the right panel of Fig.~\ref{fig:predictive_posteriors}. Yet, one can see that its value today is compatible with the standard value, and therefore the effective Newton constant (Eq.~\eqref{eq:Geff}) relevant to PN and laboratory tests respect the stringent constraints set by experiments.
The EMG model will nevertheless be tested with more accurate CMB data, as well as future measurements of large scale structure (LSS) observables from e.g. DESI, Euclid and LSST, as the matter power spectrum is different than in $\Lambda$CDM (see Ref.~\cite{Braglia:2020auw} for constraints from BOSS data on this model and forecasts for Jordan-Brans-Dicke models with future galaxy surveys in \cite{Ballardini:2019tho,Ballardini:2021evv,Euclid:2023rjj}). Our work thus motivates study with future data, as well as different form of the nonminimal coupling or the potential, that may in turn help further alleviate cosmic tensions.

\acknowledgments
The authors would like to thank Tristan Smith for very useful discussions and comments on the draft, as well as Lennart Balkenhol and Karim Benabed  for providing and helping us with the SPT likelihood for MontePython.
GFA is supported by the European Research Council (ERC) under the European Union's Horizon 2020 research and innovation programme (Grant agreement No. 864035 - Undark).  
MBa and FF acknowledge financial support from the INFN InDark initiative, from
the COSMOS network (www.cosmosnet.it) through the ASI (Italian Space Agency)
Grants 2016-24-H.0 and 2016-24-H.1-2018, as well as 2020-9-HH.0 (participation in
LiteBIRD phase A), and from the contract ASI/ INAF for the Euclid mission n.2018-23-HH.0. VP is supported by funding from the ERC under the
European Union’s HORIZON-ERC-2022 (Grant agreement No.~101076865). 
All tables and figures have been produced using the \texttt{GetDist} \cite{Lewis:2019xzd} and \texttt{fgivenx} \cite{fgivenx} packages.

\appendix

\section{Linearly perturbed Einstein equations}
\label{app:eq}

Following the conventions of Refs.~\cite{Ma:1995ey,Ballardini:2023mzm}, the linearly perturbed Einstein equations (around a FLRW background) in the synchronous gauge for a general $F(\sigma)$ are:
\ba
k^2\eta-\frac{1}{2}\mathcal{H}h'&=&-\frac{a^2 \delta\tilde{\rho}}{2 F}\\
k^2\eta'&=&\frac{a^2(\tilde{\rho}+\tilde{p})\tilde{\theta}}{2 F}\\
h''+2\mathcal{H}h'-2k^2\eta&=&-3 a^2\frac{\delta\tilde{p}}{F}\\
(h+6\eta)''+2\mathcal{H}(h+6\eta)'-2k^2\eta&=&-3 a^2\frac{(\tilde{\rho}+\tilde{p})\tilde{\sigma}}{F}
\ea

where
\ba
\label{eq:deltarhoemg}
\delta\tilde{\rho}&=&\delta\rho+\frac{\sigma'}{a^2}\delta\sigma'+V_\sigma\delta\sigma-3\frac{{\cal H}^2}{a^2}\delta F-\frac{k^2}{a^2}\delta F-\frac{3{\cal H}}{a^2}\delta F'-\frac{F' h'}{2 a^2}\\
(\tilde{\rho}+\tilde{p})\tilde{\theta}&=&(\rho+p)\theta+\frac{k^2}{a^2}\left(\sigma'\delta\sigma+\delta F'-{\cal H}\delta F\right)\\
\delta\tilde{p}&=&\delta p+\frac{\sigma'\delta\sigma'}{a^2}-V_\sigma\delta\sigma+\frac{\delta F''}{a^2}+\frac{2}{3}\frac{k^2}{a^2}\delta F+\frac{{\cal H}\delta F'}{a^2}+\frac{F' h'}{3 a^2}+(2{\cal H}'+{\cal H}^2)\frac{\delta F}{a^2}\\ \label{eq:anisotropicemg}
(\tilde{\rho}+\tilde{p})\tilde{\sigma}&=&(\rho+p)\sigma_{\rm stress}+\frac{2}{3}\frac{k^2}{a^2}\left[\delta F+\frac{F'}{k^2}\left(\frac{h'+6\eta'}{2}\right)\right].
\ea
In the equations above $'$ denotes a derivative with respect to the conformal time $a(\tau)\,\dd\tau\equiv\dd t$, $\delta\sigma$ is the perturbation to the non-minimally coupled scalar field and $ \delta F= F_\sigma\,\delta\sigma$.

\section{Semi-analytic understanding of the Sachs-Wolfe term}
\label{app:SW}
To better understand the different signatures that the EMG and RnR models leave on the CMB spectra, it is instructive to explore in detail the connection between the Weyl potential (last panel of Fig.~\ref{fig:perturbations}) and the SW effect (first panel of Fig.~\ref{fig:cmb}). Here we follow the same approach as the one described in appendix D of Ref.~\cite{Schoneberg:2023rnx}. While this approach does not provide a very accurate description of the SW temperature spectrum, it is very useful in order to gain a qualitative understanding of the underlying physics. In the line-of-sight formalism, and following the instantaneous decoupling approximation, the SW contribution to the photon temperature multipoles can be written as
\begin{equation}
\Theta_{\ell}^{\rm SW}(\tau_0,k) \simeq  \left(\frac{\delta_\gamma (\tau_{\rm dec},k)}{4}+\psi(\tau_{\rm dec},k) \right) j_{\ell}(k(\tau_0-\tau_{\rm dec})), 
\end{equation}
where $\delta_\gamma$ is the photon density perturbation and $\psi$ is the Newtonian potential accounting for gravitational redshift. Under the tight-coupling approximation, $\delta_\gamma$ can be described by the equation of a damped, driven, harmonic oscillator with the Weyl potential $\Phi \equiv (\phi+\psi)/2$ playing the role of a driving force. As first shown in Ref.~\cite{Hu:1994uz}, the solution to this equation can be written as \footnote{The solution in Eq.~\eqref{eq:delta_gamma} could in principle receive corrections due to the modification of gravity at early times. However, we expect those corrections to be very small since we are considering values $\xi \sigma_i^2 \ll 1$.   }
\begin{equation}
\frac{\delta_\gamma (\tau_{\rm dec},k)}{4} \simeq -\cos \left(k r_s(\tau_{\rm dec}) \right) -\frac{2k}{\sqrt{3}} \int_0^{\tau_{\rm dec}} d\tau'\Phi(\tau',k) \sin \left(k (r_s(\tau_{\rm dec}) -r_s(\tau') )\right) + \phi (\tau_{\rm dec},k).
\label{eq:delta_gamma}
\end{equation}

\begin{figure}[t!]
    \centering
    \includegraphics[width=0.9\columnwidth]{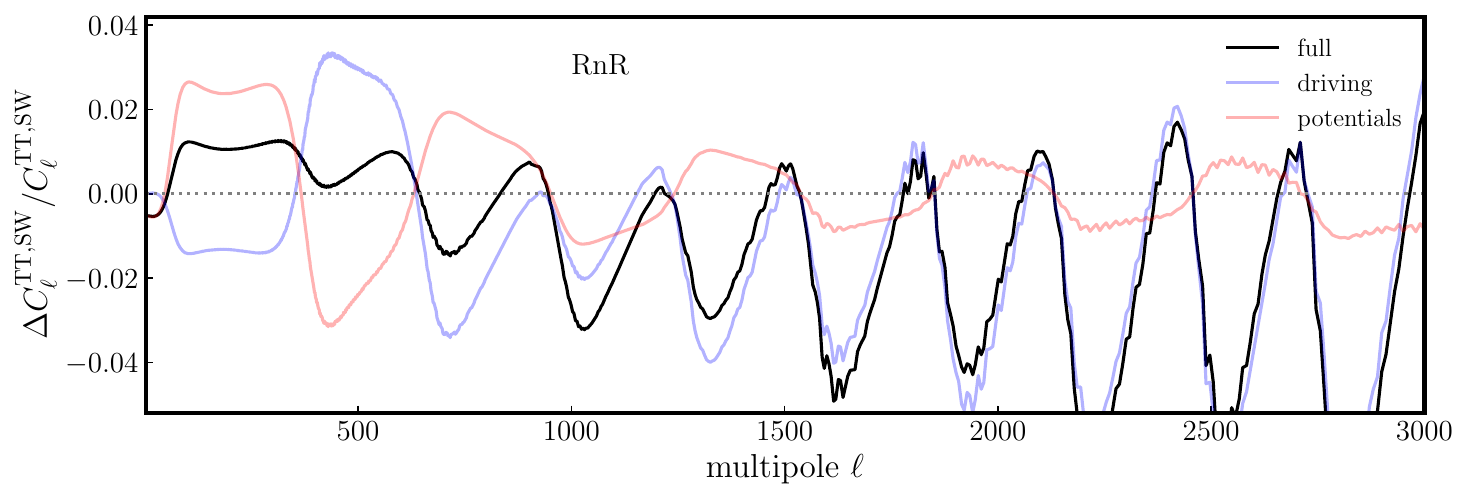}
    \includegraphics[width=0.9\columnwidth]{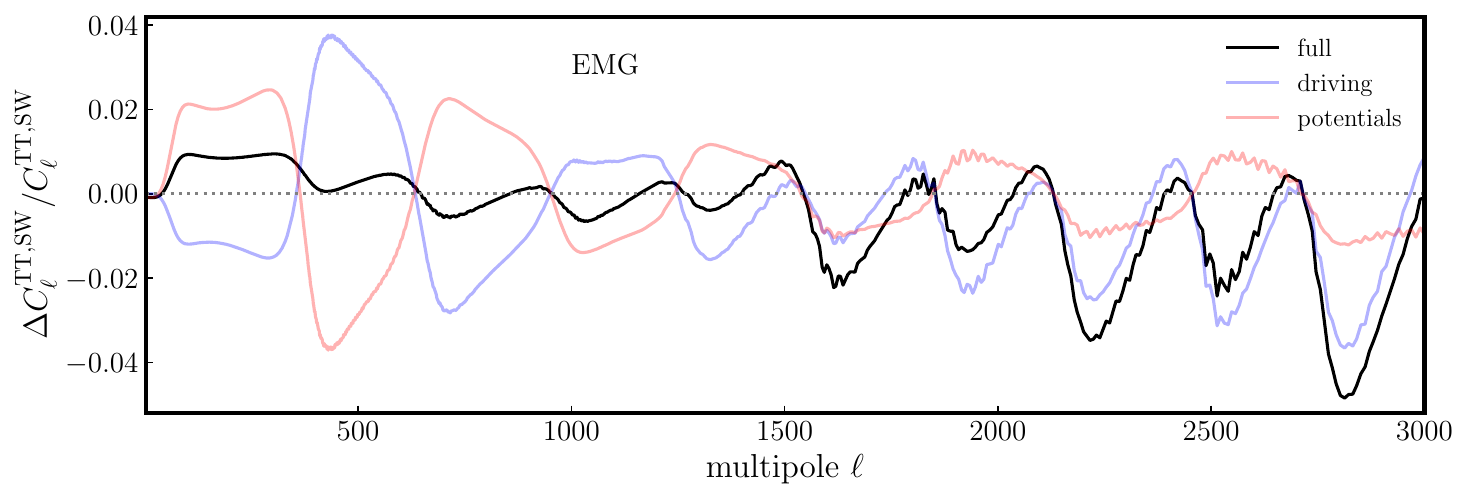}
    \caption{ The approximate fractional difference (Eq.~\eqref{eq:Delta_sw}) between the temperature power spectrum (only SW part) for both the RnR (top panel) and EMG (bottom panel). We use the same model parameters as in Fig.~\ref{fig:cmb}, except that this time we always fix the six $\Lambda$CDM parameters to the values of the EMG best-fit to {\em Planck}+ACT+SPT. The full difference (black) is the sum of the driving (blue) and potential (red) contributions. }
    \label{fig:SW_approximate}
\end{figure}

We see that this consists of several terms: a term giving free oscillations \footnote{Notice that for adiabatic initial conditions we have used the relation $ -\phi(0,k) + \delta_\gamma(0,k)/4 = -\mathcal{R}$, and we have normalized the comoving curvature perturbation, $\mathcal{R} = 1$. }, an integrated term due to the driving force of the Weyl potential, and a term due to the potential $\phi$. This solution can be refined by multiplying $\delta_\gamma$ by a damping factor $e^{-k^2/k_{\rm D}^2}$, where $k_{\rm D}$ is the diffusion wavenumber at decoupling. The temperature power spectrum is simply obtained from the photon multipoles as $ C_\ell^{\rm TT} = \int \frac{d\mathrm{ln}k }{2\pi^2}  \Theta_{\ell}^2(\tau_0,k) \mathcal{P}_{\mathcal{R}}(k)$, where $\mathcal{P}_{\mathcal{R}}(k)$ is the primordial power spectrum. Hence, for small deviations with respect to $\Lambda$CDM, we can approximate the change in the SW contribution to the temperature power spectrum as
\begin{align}
\Delta C_{\ell}^{\rm TT, SW}  \simeq  \int \frac{d\mathrm{ln}k}{\pi^2} \mathcal{P}_{\mathcal{R}}(k) \Theta_{\ell}^{\rm{SW},\Lambda \rm{CDM}}(\tau_0,k) \Delta \Theta_{\ell}^{\rm{SW}}(\tau_0,k),
\label{eq:Delta_sw}
\end{align}
where $\Theta_{\ell}^{\rm{SW},\Lambda \rm{CDM}}$ is the standard $\Lambda$CDM SW term, and $\Delta \Theta_{\ell}^{\rm{SW}}$ is given by
\begin{align}
\Delta \Theta_{\ell}^{\rm{SW}} &= \left[ -\frac{2k }{\sqrt{3}} e^{-k^2/k_{\rm D}^2} \int_0^{\tau_{\rm dec}} d\tau'\Delta \Phi(\tau',k) \sin \left(k (r_s(\tau_{\rm dec}) -r_s(\tau') )\right)  \nonumber \right.\\
&+\left. \Delta \phi (\tau_{\rm dec},k) e^{-k^2/k_{\rm D}^2}+\Delta \psi(\tau_{\rm dec},k) \right] j_{\ell}(k(\tau_0-\tau_{\rm dec})).
\end{align}
We observe that the difference in the temperature spectrum is a balance between the change in the integrated time-evolution of $\Phi$ prior to decoupling (dubbed ``driving'') 
and the change in the gravitational potentials $\phi$ and $\psi$ at decoupling (dubbed ``potentials''). Now, from last panel of Fig.~\ref{fig:perturbations}, we expect the change in the potential terms to be similar for EMG and RnR, since both yield similar values of $\phi$ and $\psi$ at decoupling. On the contrary, the effect on the driving term is vastly different for the two models. For RnR, $\Delta \Phi$ is always positive, so the contribution to the driving term is only additive; in the EMG case, the NMC leads to a negative $\Delta \Phi$ initially, thereby partially compensating the subsequent positive contribution (specially at larger $k$/larger $\ell$). We illustrate this behaviour in Fig.~\ref{fig:SW_approximate}, where we plot Eq.~\eqref{eq:Delta_sw} for EMG and RnR, considering the total expression as well as the separate contributions of the driving and potential terms. One can see that in the EMG model, the SW term receives about equal contribution from the driving and potential terms. The situation is significantly different in the RnR model: most of the contribution to the SW term is due to the driving terms, that deviate from zero in a larger way than in the EMG model at high multipoles.

We conclude that the time evolution of the gravitational potentials is crucial in order to distinguish the impact of EMG/RnR models on the CMB. While both models can produce the required dip at $\ell \sim 750$, the NMC gives enough flexibility to the EMG model to minimize the effect on the driving term and hence achieve a good fit to the CMB at $\ell \gtrsim 1000$.

\newpage
 
\section{Additional tables}
\label{app:chi2}

\begin{table*}[h!]
 \scalebox{0.8}{
 \begin{tabular}{|l|c|c|c|c|} 
    \hline  Parameter & \textit{Planck} &  \textit{Planck}+SPT &  \textit{Planck}+ACT & \textit{Planck}+ACT+SPT\\
    \hline \hline 
    $V_0$ & $2.45(2.35)^{+0.13}_{-0.38}$ 
	 & $2.48(2.29)^{+0.12}_{-0.43}$ 
	 & $2.28(2.3)^{+0.11}_{-0.13}$ 
	 & $2.277(2.206)^{+0.099}_{-0.19}$ 
	 \\
    $ \sigma_i$ & $0.33(0.36)^{+0.12}_{-0.15}$ 
	 & $0.30(0.38)\pm 0.12$ 
	 & $0.392(0.415)\pm 0.094$ 
	 & $0.37(0.38)^{+0.14}_{-0.12}$ 
	 \\
    $\xi \sigma_i^2$ 	 & $<0.054 (0.036)$ 
	 & $<0.04(0.027) $ 
	 & $0.041(0.052)\pm 0.011$ 
	 & $0.033(0.037)\pm 0.014$ 
	 \\
    \hline  \rule{0pt}{3ex}
    $h$ & $0.6858(0.6858)^{+0.0083}_{-0.013}$ 
	 & $0.6839(0.6866)^{+0.0075}_{-0.0095}$ 
	 & $0.6888(0.6887)^{+0.0086}_{-0.010}$ 
	 & $0.6890(0.6885)^{+0.0092}_{-0.011}$ 
	 \\
    $100\omega_b$ 	 & $2.248(2.245)\pm 0.019$ 
	 & $2.243(2.244)\pm 0.017$ 
	 & $2.234(2.231)\pm 0.018$ 
	 & $2.235(2.229)\pm 0.016$ 
	 \\
    $\omega_{\rm cdm}$ 	 & $0.1251(0.1265)^{+0.0022}_{-0.0048}$ 
	 & $0.1238(0.1257)^{+0.0023}_{-0.0037}$ 
	 & $0.1280(0.1299)^{+0.0031}_{-0.0037}$ 
	 & $0.1264(0.1264)^{+0.0031}_{-0.0045}$ 
	 \\
    $\mathrm{ln}(10^{10} A_s)$ 	 & $3.054(3.055)^{+0.016}_{-0.020}$ 
	 & $3.046(3.049)\pm 0.017$ 
	 & $3.059(3.061)\pm 0.016$ 
	 & $3.052(3.051)\pm 0.016$ 
	 \\
    $n_s$ 	 & $0.9700(0.9708)^{+0.0060}_{-0.0067}$ 
	 & $0.9691(0.9692)\pm 0.0054$ 
	 & $0.9706(0.9731)^{+0.0051}_{-0.0065}$ 
	 & $0.9701(0.9684)\pm 0.0055$ 
	 \\
    $\tau_{\rm reio}$ 	 & $0.0547(0.0526)^{+0.0072}_{-0.0086}$ 
	 & $0.0521(0.052)\pm 0.0080$ 
	 & $0.0514(0.0502)\pm 0.0074$ 
	 & $0.0502(0.0504)\pm 0.0079$ 
	 \\
    \hline \rule{0pt}{3ex}
    $f_{\rm scf}(z_c)$ 	 & $0.070(0.099)^{+0.028}_{-0.055}$ 
	 & $0.056(0.0)^{+0.022}_{-0.044}$ 
	 & $0.117(0.139)\pm 0.030$ 
	 & $0.0995(0.1064)\pm 0.039$ 
	 \\
    $\log_{10}(z_c)$ 	 & $3.626(3.614)^{+0.053}_{-0.12}$ 
	 & $3.621(3.587)^{+0.058}_{-0.15}$ 
	 & $3.589(3.617)^{+0.046}_{-0.034}$ 
	 & $3.564(3.551)\pm 0.045$ \\
    $S_8$ 	 & $0.838(0.843)^{+0.016}_{-0.019}$ 
	 & $0.831(0.835)\pm 0.016$ 
	 & $0.847(0.857)\pm 0.018$ 
	 & $0.836(0.833)\pm 0.016$ 
	 \\
    $\Omega_m$ 	 & $0.3152(0.318)^{+0.0080}_{-0.0091}$ 
	 & $0.3141(0.3157)\pm 0.0081$ 
	 & $0.3184(0.3223)\pm 0.0086$ 
	 & $0.3147(0.3149)\pm 0.0080$ 
	 \\
    Age [Gyrs] 	 & $13.56(13.52)^{+0.20}_{-0.098}$ 
	 & $13.61(13.53)^{+0.16}_{-0.085}$ 
	 & $13.46(13.41)^{+0.16}_{-0.12}$ 
	 & $13.50(13.5)^{+0.19}_{-0.14}$ 
	 \\
    \hline
\end{tabular} 
}
\caption{The mean (best-fit) $\pm 1\sigma$ errors (optionally 95\% C.L. upper limit)  of the cosmological parameters reconstructed from analyses of various data sets (see column title) in the EMG model. For convenience, we repeat the constraints of the $\emph{Planck}$+ACT+SPT analysis of the EMG model in the right column of Table \ref{tab:CMB_2} and the left column of Table \ref{tab:CMB_3}. }
\label{tab:CMB}
\end{table*}

\begin{table*}[h!]
\centering
 \scalebox{0.8}{
 \begin{tabular}{|l|c|c|c|} 
    \hline  Parameter & $\Lambda$CDM & RnR  & EMG \\
    \hline \hline 
$V_0$ & --
	 & $1.85(1.64)^{+0.56}_{-0.94}$ 
      & $2.277(2.206)^{+0.099}_{-0.19}$ 
	 \\
$\sigma_i$ & --
	 & $0.186(0.222)^{+0.068}_{-0.13}$ 
	 & $0.37(0.38)^{+0.14}_{-0.12}$ 
	 \\
$\xi\sigma_i^2$ & --
      & --
      & $0.033(0.037)\pm 0.014$ 
      \\
\hline  \rule{0pt}{3ex}
    
$h$
	 & $0.6743(0.6747)\pm 0.0050$ 
	 & $0.6771(0.6768)^{+0.0053}_{-0.0068}$ 
	 & $0.6890(0.6885)^{+0.0092}_{-0.011}$ 
	 \\
$100 \omega_{b }$
	 & $2.234(2.236)\pm 0.012$ 
	 & $2.232(2.223)^{+0.016}_{-0.013}$ 
	 & $2.235(2.229)\pm 0.016$ 
	 \\
$\omega_{\rm cdm }$
	 & $0.1199(0.1199)\pm 0.0012$ 
	 & $0.1206(0.1209)^{+0.0013}_{-0.0017}$ 
	 & $0.1264(0.1264)^{+0.0031}_{-0.0045}$ 
	 \\
$\mathrm{ln}(10^{10} A_s)$
	 & $3.046(3.048)\pm 0.015$ 
	 & $3.047(3.044)\pm 0.014$ 
	 & $3.052(3.051)\pm 0.016$ 
	 \\
$n_{s }$
	 & $0.9686(0.9696)\pm 0.0038$ 
	 & $0.9689(0.9662)\pm 0.0046$ 
	 & $0.9701(0.9684)\pm 0.0055$ 
	 \\
$\tau_{\rm reio}$ 
      & $0.0515(0.0526)\pm 0.0076$ 
	 & $0.0521(0.0509)\pm 0.0072$ 
	 & $0.0502(0.0504)\pm 0.0079$ 
	 \\
\hline \rule{0pt}{3ex}

$f_{\rm scf}(z_c)$ 
& --
& $<0.034(0.011)$
& $0.0995(0.1064)\pm 0.039$ 
\\

$\log_{10}(z_c)$ 
& -- 
 & $3.11(3.09)^{+0.49}_{-0.41}$ 
 & $3.564(3.551)\pm 0.045$ 
	 \\
$S_8$
	 & $0.832(0.833)\pm 0.014$ 
	 & $0.830(0.83)\pm 0.014$ 
  & $0.836(0.833)\pm 0.016$ 
	 \\
$\Omega_{m }$
	 & $0.3144(0.3139)\pm 0.0070$ 
	 & $0.3133(0.3139)\pm 0.0072$ 
	 & $0.3147(0.3149)\pm 0.0080$ 
	 \\
Age [Gyrs]
	 & $13.793(13.791)\pm 0.019$ 
	 & $13.751(13.749)^{+0.066}_{-0.020}$ 
	 & $13.50(13.5)^{+0.19}_{-0.14}$ 
	 \\
    \hline
\end{tabular} 
}
\caption{The mean (best-fit) $\pm 1\sigma$ errors (optionally 95\% C.L. upper limit) of the cosmological parameters reconstructed from analyses of the data set \emph{Planck}+ACT+SPT in different models (see column title).}
\label{tab:CMB_2}
\end{table*}

\begin{table*}[h!]
\centering
 \scalebox{0.8}{
 \begin{tabular}{|l|c|c|c|} 
    \hline  Parameter & \textit{Planck}+ACT+SPT  & \textit{Planck}+ACT+SPT+Ext   & \textit{Planck}+ACT+SPT+Ext+$M_b$ \\
    \hline \hline 
$V_0$
      & $2.277(2.206)^{+0.099}_{-0.19}$ 
	 & $2.22(2.2)\pm 0.19$ 
	 & $2.065(2.032)^{+0.066}_{-0.078}$ 
	 \\
$\sigma_i$
	 & $0.37(0.38)^{+0.14}_{-0.12}$ 
	 & $0.37(0.39)\pm 0.13$ 
	 & $0.632(0.643)^{+0.071}_{-0.054}$ 
	 \\
$\xi\sigma_i^2$ 
      & $0.033(0.037)\pm 0.014$ 
	 & $0.032(0.038)^{+0.016}_{-0.011}$ 
	 & $0.054(0.053)\pm 0.010$ 
	 \\ 
\hline  \rule{0pt}{3ex}
$h$
	 & $0.6890(0.6885)^{+0.0092}_{-0.011}$ 
	 & $0.6892(0.6891)^{+0.0075}_{-0.012}$ 
	 & $0.7155(0.7165)\pm 0.0074$ 
	 \\
$100\omega_{b }$
	 & $2.235(2.229)\pm 0.016$ 
	 & $2.235(2.229)\pm 0.015$ 
	 & $2.255(2.253)\pm 0.014$ 
	 \\
$\omega_{\rm cdm }$
	 & $0.1264(0.1264)^{+0.0031}_{-0.0045}$ 
	 & $0.1264(0.1267)^{+0.0030}_{-0.0045}$ 
	 & $0.1355(0.1354)\pm 0.0035$ 
	 \\
$\mathrm{ln}(10^{10} A_s)$
	 & $3.052(3.051)\pm 0.016$ 
	 & $3.054(3.053)\pm 0.014$ 
	 & $3.069(3.069)\pm 0.014$ 
	 \\
$n_{s }$
	 & $0.9701(0.9684)\pm 0.0055$ 
	 & $0.9693(0.968)\pm 0.0052$ 
	 & $0.9793(0.9783)\pm 0.0050$ 
	 \\
$\tau_{\rm reio}$
	 & $0.0502(0.0504)\pm 0.0079$ 
	 & $0.0514(0.0505)\pm 0.0071$ 
	 & $0.0530(0.0537)\pm 0.0070$ 
	 \\
\hline \rule{0pt}{3ex}

$f_{\rm scf}(z_c)$ 
      & $0.0995(0.1064)\pm 0.039$ 
	 & $0.098(0.0)^{+0.043}_{-0.036}$ 
	 & $0.180(0.181)\pm 0.025$ 
	 \\
$\log_{10}(z_c)$ 
   & $3.564(3.551)\pm 0.045$ 
	 & $3.532(3.551)^{+0.073}_{-0.019}$ 
	 & $3.570(3.557)\pm 0.034$ 
	 \\
$S_8$
  & $0.836(0.833)\pm 0.016$ 
	 & $0.835(0.834)\pm 0.011$ 
	 & $0.837(0.834)\pm 0.012$ 
	 \\
$\Omega_{m }$
	 & $0.3147(0.3149)\pm 0.0080$ 
	 & $0.3144(0.3151)\pm 0.0053$ 
	 & $0.3099(0.3088)\pm 0.0051$ 
	 \\
Age [Gyrs]
	 & $13.50(13.5)^{+0.19}_{-0.14}$ 
	 & $13.50(13.49)^{+0.21}_{-0.13}$ 
	 & $13.05(13.05)\pm 0.13$ 
	 \\
    \hline
\end{tabular} 
}
\caption{The mean (best-fit) $\pm 1\sigma$ errors of the cosmological parameters reconstructed from analyses of various data sets (see column title) in the EMG model. Here `Ext' refers to the data combination `\textit{Planck}~lensing+BAO/$f\sigma_8$+PanPlus'. }
\label{tab:CMB_3}
\end{table*}

\begin{table*}[t]
    \centering
\scalebox{0.9}{ 
    \begin{tabular}{|c|ccc c c c c c| c| c| c c}
    \hline     Model & low$-\ell$  & high$-\ell$ &  SPT & ACT   & BAO/$f\sigma_8$ & Pan+ & Lens & $M_b$  & $\Delta \chi^2$tot & $\Delta  \mathrm{AIC}$ \\ \hline \hline
         EMG &  $-0.11$ &  $-3.86$  & -- & -- & -- & --  & -- & -- & $-3.97$ & $+2.03$  \\ 
         EMG & $+0.26$ & $-2.97$  & $-0.42$ & -- & --  & --  & -- & -- & $-3.13$ & $+2.87$   \\
         EMG & $+0.34$ & $-4.37$  & --  & $-6.21$ & -- & --  & --  & --& $-10.24$ & $-4.24$    \\
         EMG & $+1.24$ & $-3.53$ & $+0.03$  & $-5.86$ &  -- &  -- & --  & -- & $-8.12$ & $-2.12$     \\
         EMG & $+1.5$ & $-3.58$ & $-0.01$  & $-6.21$ & $+0.42$   & $-0.08$  & $-0.23$  &  -- & $-8.19$ & $-2.19$     \\
         EMG & $+0.77$ & $+1.29$ & $-1.36$  & $-7.43$ & $+0.30$ & $-1.16$ & $+0.08$  & $-30.69$ & $-38.2$ & $-32.2$      \\
        \hline \rule{0pt}{3ex}
         RnR  & $+0.49$ & $+1.5$  & $-0.10$  & $-2.78$ & -- & -- & -- &  -- & $-0.89$ &$+3.11$ \\  \hline
    \end{tabular}
}
    \caption{Contribution of each likelihood to the best-fit  $\chi^2 = -2\ln{\mathcal{L}}$ of EMG and RnR models to different datasets,
relative to the best-fit $\chi^2$ of $\Lambda$CDM to the same datasets. In the last column we report the corresponding $\Delta$AIC. }
    \label{tab:chi2}
\end{table*}

\begin{table*}[t]
\centering
\begin{tabular}{|c|c|c|}
\hline Model & Gaussian tension  & $Q_{\mathrm{DMAP}}$ tension   \\ 
\hline \hline
$\Lambda$CDM & $6.37\sigma$ & $6.36\sigma$ \\ \hline
EMG                   & $3.82\sigma$ &  $3.25\sigma$  \\ \hline
    \end{tabular}
\caption{Different tension metrics between the datasets \emph{Planck}+ACT+SPT+Ext and S$H_0$ES for $\Lambda$CDM and EMG models, expressed in terms of the absolute SN Ia magnitude $M_b$. Here `Ext' refers to the data combination `\textit{Planck}~lensing+BAO/$f\sigma_8$+PanPlus'. }
    \label{tab:tension_metrics}
\end{table*}

\newpage

\bibliographystyle{JHEP}
\bibliography{EMG}

\end{document}